\documentclass[aps, a4paper, twocolumn, nofootinbib, showpacs, amsfonts]{revtex4}

\usepackage[utf8]{inputenc}
\usepackage{amssymb}

\usepackage{graphicx}
\usepackage{amsmath}
\usepackage{bm}
\usepackage{xcolor}
\usepackage{tikz}
\usepackage{mathrsfs}
\usetikzlibrary{arrows}
\usepackage{cancel}
\usepackage[normalem]{ulem}

\newcommand{\bea}{\begin{eqnarray}}
\newcommand{\eea}{\end{eqnarray}}
\newcommand{\be}{\begin{equation}}
\newcommand{\ee}{\end{equation}}

\begin{document}

\title{Observational properties of regular black holes in Asymptotic Safety}

\author{Abdybek Urmanov}
\email{abdybek.urmanov@nu.edu.kz}
\affiliation{Department of Physics, Nazarbayev University, Kabanbay Batyr 53, 010000 Astana, Kazakhstan}

\author{Hrishikesh Chakrabarty}
\email{hrishikesh.chakrabarty@nu.edu.kz}
\affiliation{Department of Physics, Nazarbayev University, Kabanbay Batyr 53, 010000 Astana, Kazakhstan}

\author{Daniele Malafarina}
\email{daniele.malafarina@nu.edu.kz}
\affiliation{Department of Physics, Nazarbayev University, Kabanbay Batyr 53, 010000 Astana, Kazakhstan}
    
\begin{abstract}
We consider the observational properties of a spherically symmetric, static regular black hole within the framework of asymptotic safety (AS) as proposed by Bonanno et al. The metric resembles the Schwarzschild solution in the classical limit.  
The departure from Schwarzschild at small scales is controlled by a single free parameter related to the ultraviolet (UV) cutoff of the theory.
We investigated null and time-like geodesics around the AS metric, including circular orbits, photon rings 
and lensing effects. In particular we focused on the optical properties of thin accretion disks in the equatorial plane of the object
and compared them with those of accretion disks in the Schwarzschild metric. We found that the radiation flux, luminosity, and efficiency of the accretion disk increase with the value of the free parameter. Using a spacetime generic open-source relativistic ray-tracing code, we simulate the K$\alpha$ iron line profiles emitted by the disk and analyze their deviation from that of the Schwarzschild geometry.

\end{abstract}

\maketitle

\section{Introduction}\label{sec1}
One of the most fascinating predictions of General Relativity (GR) is the existence of black holes. 
In recent years this prediction has been validated by recent astrophysical discoveries, including the detection of gravitational waves from binary mergers by LIGO \cite{PhysRevLett.116.061102} and the direct imaging of supermassive black holes at the centers of the M87 and Milky Way galaxies by the Event Horizon Telescope collaboration \cite{EventHorizonTelescope:2019dse,EventHorizonTelescope:2019ths}. 

In the framework of GR, the defining characteristic of a black hole is the existence of spacetime singularity at its center, which is hidden from external observers by the event horizon \cite{Hawking:1970zqf}. Since the causal nature of the theory breaks down approaching spacetime singularities,  
they are generally regarded as an artifact arising due to the limits of validity of GR \cite{Senovilla:1998oua}. Consequently, in recent years models that extend beyond GR and resolve singularities have gained significant interest as probes into the properties that a new theory of gravity should posses \cite{Cardoso:2019rvt}. From such approaches a variety of exotic compact objects such as regular black holes \cite{Bambi:2023try,Lan_2023}, gravastars \cite{Visser:2003ge}, boson stars\cite{Olivares:2018abq}, wormholes \cite{Visser:1989kh}, quantum coherent black holes\cite{universe7120478,Urmanov:2024qai}, etc., have been proposed. 
Interestingly, the previously mentioned recent achievements in precision observations of the near horizon region have opened the door to the possibility of testing and potentially rule out such proposals.
This has led to extensive research focused on analyzing the electromagnetic signatures of the surroundings of various candidates \cite{Bambi:2015kza}, including lensing effects \cite{Chakrabarty:2022fbd}, accretion disk luminosity \cite{Joshi_2013}, X-ray data \cite{Bambi:2013qj}, and others \cite{Bambi_2012a}. 

The first proposal for a regular black hole (RBH) solution was made by Bardeen 
in 1968 \cite{1968qtr..conf...87B}. Since then numerous models have been developed. The simplest approach to constructing RBHs involves modifying the Misner-Sharp mass to ensure it rapidly approaches zero near the center in such a way that curvature invariants remain finite. The behavior of the mass function near the center may be dictated by considerations on the expected behavior of gravity at large curvatures. 
Notable examples of this approach include the Poisson-Israel model \cite{1988CQGra...5L.201P}, the Dymnikova regular black hole \cite{Dymnikova:1992ux} and the Hayward metric \cite{Hayward_2006}. Other RBHs have been obtained from coupling GR to a theory of non-linear electrodynamics \cite{Ayon-Beato:1998hmi, Bronnikov:2000vy}, in conformal gravity \cite{Bambi:2016wdn,Chakrabarty:2017ysw,Jusufi:2019caq} etc.
Ideally one would want the RBH solution to arise dynamically from gravitational collapse in a modified theory with a well posed action. This is the case for example of the Hayward and Bardeen black holes, which may also be obtained from GR coupled to a theory of non-linear electrodynamics \cite{Fan:2016hvf, Bronnikov:2017tnz, Toshmatov:2018cks, Malafarina:2022oka}.
Regarding the search for a theory of quantum gravity, we shall recall a few notable attempts to singularity resolutions such as the null shell collapse by Frolov and Vilkovisky \cite{Frolov:1981mz}, the first RBH in AS by 
Bonanno and Reuter \cite{Bonanno_2000}
and the Loop Quantum Gravity dust collapse model in \cite{Bambi:2013caa}.
For a review of non-singular collapse models see \cite{Malafarina:2017csn} while for recent survey of the state of the art in research on RBHs and black hole mimickers see \cite{Carballo-Rubio:2025fnc} and references therein.

A recent proposals for a new class of regular black holes in Asymptotic Safety (ASBH) was presented by Bonanno et al. \cite{PhysRevLett.132.031401}. In this work, the authors derived an explicit metric that describes the exterior of a collapsing dust ball within the framework of Asymptotically Safe gravity. Their approach implies a formulation of the antiscreening behavior of gravity in ultra-Planckian energy domains, incorporating a multiplicative coupling with the matter Lagrangian 
based on an initial idea by Markov and Mukhanov \cite{1984JETPL..40.1043M}. 
The main difference between the theory from which the RBH in \cite{PhysRevLett.132.031401} is obtained and the one used for \cite{Bonanno_2000} resides in the addition of the coupling function in the matter Lagrangian, which effectively allows to identify the cutoff of the theory with a critical value of the energy density of the source.

In this article, we study in detail the motion of both massive and massless particles around the ASBH 
and investigate its electromagnetic observational properties. Investigating the surroundings of regular black holes, such as accretion disks, could reveal the effect of quantum corrections, which would help distinguish them from the standard model. Several works are already focused on analyzing shadow and quasinormal modes \cite{Stashko_2024,Sanchez:2024sdm}. 

The paper is organized as follows: In section \ref{sec2}, we review the spacetime metric and its basic properties. Sections \ref{sec3} and \ref{sec4} focus on the motion of massive and massless test particles. Additionally, we analyze observational properties; in section \ref{sec3}, we investigate gravitational lensing in the geometry of the AS, while in section \ref{sec4}, we examine the luminosity of thin accretion disks. In section \ref{sec5}, we use an open source and generic general relativistic ray-tracing code to simulate the K$\alpha$ iron line profile. Finally, in section \ref{sec6}, we discuss obtained results, possible implications, and outlook. Throughout this article, we consider geometric units $c = G =1$,
and we employ the mostly positive metric signature $\{-,+,+,+\}$.

\section{The spacetime}\label{sec2}

The line element of the static, spherically symmetric, regular black hole in asymptotic safety (ASBH) \cite{PhysRevLett.132.031401} in Schwarzschild coordinates $\{t,r,\theta,\phi\}$ is given by
\begin{equation}
    ds^2=-f(r)dt^2+\frac{1}{f(r)}dr^2+r^2d\Omega^2,
    \label{metric}
\end{equation}
where $d\Omega^2=d\theta^2+\sin^2{\theta}d\phi^2$, $f(r)=1-2M(r)/r$ and $M(r)$ is the Misner-Sharp mass function that captures the deviation from Schwarzschild due to asymptotic safety. The function $M(r)$ is given by 
\begin{equation}
    M(r)=\frac{r^3}{6\xi}\log\Big(1+\frac{6M_0\xi}{r^3}\Big),
    \label{mass_function}
\end{equation}
where $M_0$ is the mass parameter and $\xi$ is a free dimensional parameter that sets the scale for Ultra Violet (UV) cutoff of Asymptotic Safety  \cite{PhysRevLett.132.031401}. 
For $\xi\rightarrow0$ or similarly $r\rightarrow\infty$, $M(r)$ approximates to
\begin{equation}
    M(r)\simeq M_0-\frac{3M_0^2\xi}{r^3}+O(\xi^2),
\end{equation}
and the metric reduces to Schwarzschild with constant mass $M_0$. For regimes near the center $(r\rightarrow 0)$, $M(r)$ approximates to
\begin{equation}
    M(r)\simeq \frac{r^3}{6\xi}\log\left(\frac{6M_0\xi}{r^3}\right)+O(r^6).
\end{equation}

The solution presented in \cite{PhysRevLett.132.031401} was obtained from dynamical collapse in AS in a theory with non minimal coupling of gravity to matter as proposed in \cite{1984JETPL..40.1043M}. In this theory the action takes the form
\be 
\mathcal{A} = \int d^4x \sqrt{-g} \left( \frac{R}{8\pi G_N} + 2\chi (\rho) \mathcal{L}_{\rm m} \right),
\ee 
with the coupling function $\chi$ depending on the energy density $\rho$. 
The modified field equations can then be written as
\begin{equation}
R_{\mu\nu} - \frac{1}{2}g_{\mu\nu}R = 
   8\pi G(\rho)T_{\mu\nu} - \Lambda(\rho)g_{\mu\nu},
\end{equation}
where the variable coupling $G$ and $\Lambda$ are defined as
\begin{equation}
    G(\rho) = G_N(\chi \rho)_{,\rho} \; \; \text{and} \; \; \Lambda(\rho) = - 8\pi G_N \rho^2 \chi_{,\rho} .
\end{equation}
The model is fully determined once the running coupling $G$ is presecribed from AS. The black hole solution was subsequently obtained from dust collapse in a manner equivalent to the classical Oppenheimer-Snyder-Datt model in GR \cite{1939PhRv...56..455O,1938ZPhy..108..314D}.
This approach differs from the original RBH solution of Bonanno and Reuter \cite{Bonanno_2000} in which the BH solution was derived solely from the running gravitational coupling obtained via the renormalization group with the ultraviolet cutoff determined by a given length scale.

It is important to note that the ASBH given by Eq. \eqref{mass_function}
is not regular everywhere, as one can easily find that the Kretschmann Scalar diverges for $r\rightarrow 0$. In fact we have
\begin{align}
    K &= R_{\alpha\beta\mu\gamma} R^{\alpha\beta\mu\gamma} = \frac{4}{r^6} \Big( 12M^2 - 16M M' r + \nonumber\\ &+8(M')^2 r^2 +  4M'' M r^2 - 4M'' M' r^3 + (M'')^2 r^4 \Big),
\end{align}
where prime ($'$) denotes a derivative with respect to radial coordinate $r$.
However, as was shown in \cite{PhysRevLett.132.031401}, a collapsing matter cloud in the ASBH model approaches the singularity in an infinite amount of time, as opposed to what happens in the Oppenheimer-Snyder-Datt \cite{1939PhRv...56..455O,1938ZPhy..108..314D} collapse model where the singularity is achieved in a finite co-moving time.
Of course for constant $M$, we have $M'=M''=0$, and $K$ reduces to the Kretschmann scalar for Schwarzschild, i.e. $K=48M^2/r^6$.

\begin{figure}[tt]
   \begin{center}
        \includegraphics[width=9cm]{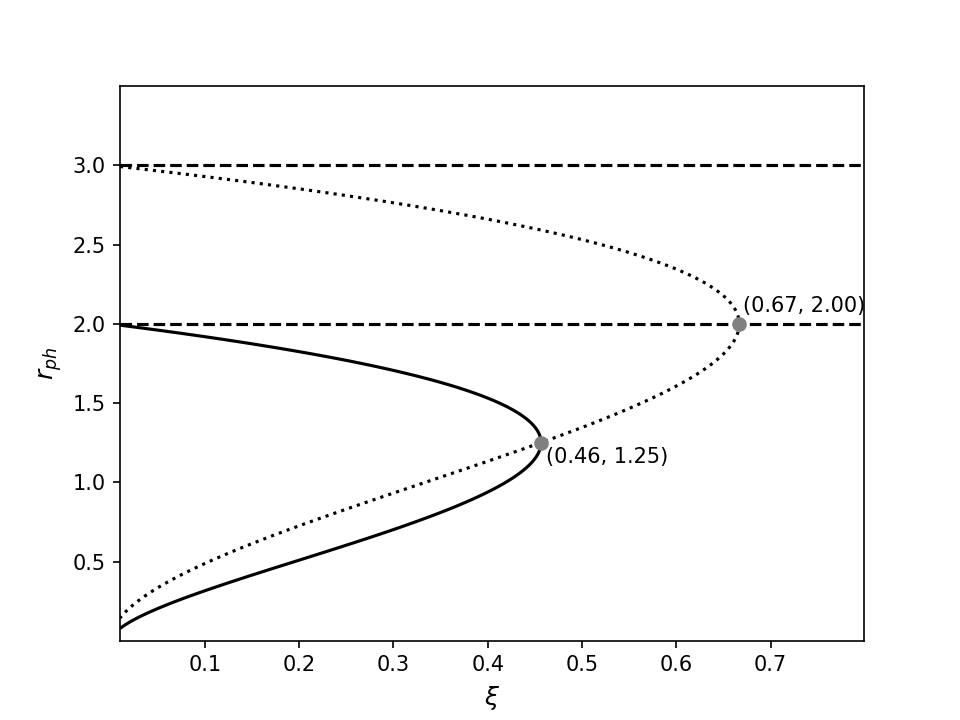}    
  \end{center}
    \caption{ The solid black line shows the horizon radius $r_h$ of the ASBH as a function of $\xi$, where for simplicity we have set $M_0=1$ (which is equivalent to a rescaling of the coordinates $r\rightarrow M_0 r  $ and $t\rightarrow t/M_0$ and $\xi\rightarrow M_0^2\xi$). For comparison, the lower horizontal dashed line shows the Schwarzschild horizon $r_{\rm Sch}=2M_0$. The gray dot on the solid line indicates the critical value $\xi_{\rm crit}\simeq 0.46$ for which the ASBH has only one horizon at $r_h=r_{\rm crit}\simeq 1.25$.
    The dotted line represents the location of the ASBH photon ring as a function of $\xi$  
    while the upper horizontal dashed line illustrates the Schwarzschild photon ring.  
    The gray dot on the dotted line highlights the limiting value of $\xi$ for which the AS geometry has only one photon ring.  
    }\label{fig:r_ph}
\end{figure}

It can also be shown that the Arnowitt-Deser-Misner (ADM) mass of the ASBH is equal to $M_0$. 
The horizon radius $r_h$ of ASBH is obtained by solving $f(r)=0$ and is shown in Fig.~\ref{fig:r_ph} as a function of the free parameter $\xi$. 
Notice that in the figures we apply a rescaling of the coordinates $r\rightarrow M_0 r  $ and $t\rightarrow t/M_0$ and the parameter $\xi\rightarrow M_0^2\xi$.
The ASBH has two horizons, an inner Cauchy horizon and an outer event horizon for values of $\xi< \xi_{\rm crit}$ and hence it is a black hole solution.
However, there is only one event horizon at the critical value of parameter $\xi_{\rm crit}\approx0.46 M_0^2$, for which the corresponding radius is $r_{\rm crit}\approx1.25 M_0$. 
For $\xi > \xi_{\rm crit}$ the eternally collapsing object is never covered by a horizon becoming what is usually referred to as a horizonless compact object.
Notice that the outer horizon reduces to Schwarzschild event horizon $r_{\rm Sch}=2M_0$ as $\xi\rightarrow 0$ and the radius of the inner horizon becomes zero. 

Since horizons cannot be directly observed we must rely on the surroundings of the ASBH in order to obtain some insights on how it may be observationally distinguished from a Schwarzschild BH of the same mass. 
Therefore, quantities that characterize the structure of the surrounding environment, such as the radius of the photon sphere or the innermost stable circular orbit, are of great importance.

\section{Motion of massless particles}\label{sec3}

In this section, we investigate null geodesics around the ASBH. 
The geometry is static and spherical symmetric and therefore, two killing vectors exist associated with two conserved quantities for particle motion: the energy $p_t=E$ and angular momentum $p_\phi=L$.
Because of spherical symmetry we can consider geodesics only in the equatorial plane ($\theta=\pi/2$) without loss of generality. Then, the conserved quantities can be identified from
\begin{align}
    &\dot{t}=p^t=\Big(1-\frac{2M}{r}\Big)^{-1}E,
    \label{t_dot} \\
    &\dot{\phi}=p^{\phi}=r^{-2}L,
    \label{phi_dot}
\end{align}
where the overhead dot represents a derivative with respect to the geodesic affine parameter. The radial equation of motion is derived using the normalization condition for massless particles $g_{\mu\nu}\dot{x}^{\mu}\dot{x}^{\nu}=0$ and it is given by,
\begin{equation}
    \dot{r}^2=E^2-V_{\rm eff} ,
    \label{r_dot}
\end{equation}
where $V_{\rm eff}$ is the effective potential for ASBH and it is given by, 
\begin{equation}
    V_{\rm eff}=\Big(1-\frac{2M(r)}{r}\Big)\frac{L^2}{r^2}.
    \label{veff_ph}
\end{equation}
\begin{figure}[tt]
   \begin{center}
        \includegraphics[width=9cm]{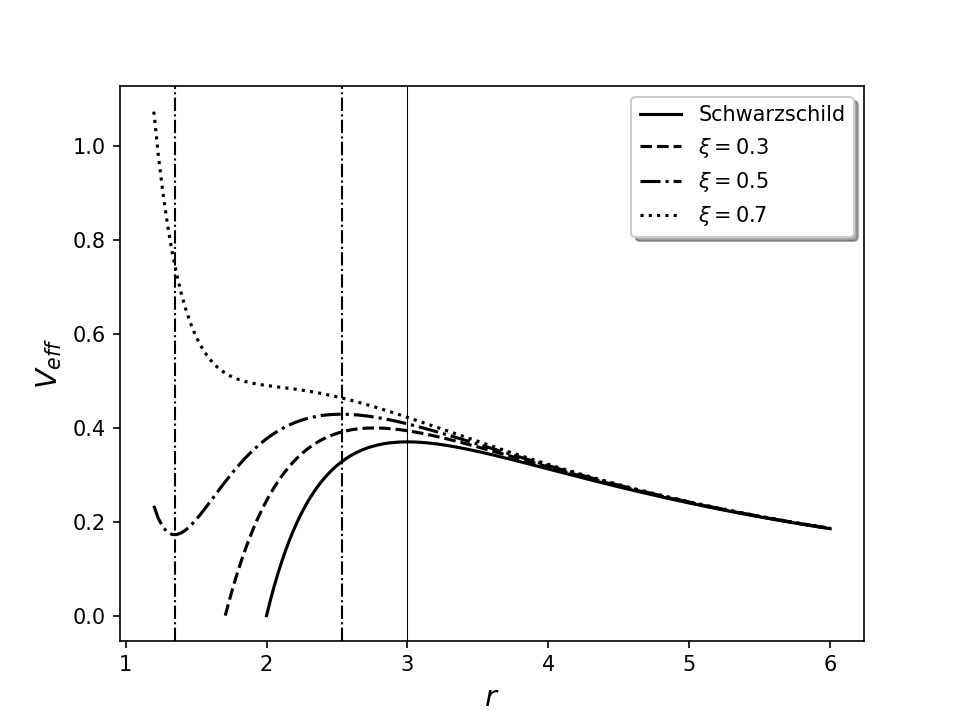}    
  \end{center}
    \caption{The effective potential for photons in the equatorial plane of the ASBH for three values of $\xi$ as a function of the radial coordinate in units of $M_0$ with the angular momentum fixed at $L=10$. The solid line represents the effective potential for the Schwarzschild metric, and the vertical gray line shows the position of the photon ring for the Schwarzschild black hole. The vertical dash-dotted lines represent the positions of photon rings for $\xi=0.5$.} \label{fig:AS_Veff_ph}
\end{figure}
Fig.~\ref{fig:AS_Veff_ph} shows the effective potential for null geodesics (\ref{veff_ph}) as a function of the radial coordinate for different values of $\xi$ compared to the Schwarzschild geometry. At large distances from the compact object, the effective potential mimics the behavior of Schwarzschild's potential regardless of the value of $\xi$. However, at short distances it deviates 
and the difference becomes visibly larger for bigger values of $\xi$. 
Most notably, the radius corresponding to the peak of the effective potential, which defines the location of the unstable photon orbit, decreases as $\xi$ increases. 

To obtain the location of the photon ring, we impose circular orbit condition on massless particles  $V_{\rm eff}'=0$, which is equivalent to $M'r+r=3M$ and results in a simple cubic equation
\begin{equation}
    r^3-3M_0r^2+6M_0\xi=0.
\end{equation}

For $\xi\rightarrow0$ we immediately recover the Schwarzschild photon ring, $r_{ph}\rightarrow3M_0$.
For $\xi/M_0^2<2/3$ we have three roots $\{R_1,R_2,R_3\}$, two of them positive and one negative. Neglecting the negative root, 
the outer ($R_1$) and inner ($R_2$) photon rings are given by,  
\begin{align}
    R_1 &= -2M_0\cos\left(\frac{\Theta+2\pi}{3}\right) + M_0, \\
    R_2 &= -2M_0\cos\left(\frac{\Theta-2\pi}{3}\right) + M_0,
\end{align}
where 
\begin{equation}
    \Theta=\arccos\left(\frac{3\xi-M_0^2}{M_0^2}\right).
\end{equation}
The ASBH photon rings structure is 
richer than the Schwarzschild one. For $0<\xi/M_0^2\lesssim 0.46$, the spacetime has both an outer and an inner horizon and the inner photon ring radius is not physical (as it is inside the outer horizon as can be seen from Fig.~\ref{fig:r_ph}), thus the ASBH has one unstable photon orbit similarly to Schwarzschild. However, for $0.46\lessapprox\xi/M_0^2 \le2/3$, the geometry has no horizons, thus describing a horizonless compact object with two photon rings. Most interestingly the inner photon radius is stable in this case.
At the critical value $\xi/M_0^2=2/3$ the photon ring radius becomes equal to the Schwarzschild radius.
Finally in the regime $\xi/M_0^2>2/3$, circular null orbits can not be formed and the exotic compact object exhibits a Newtonian behavior.  
In Fig.~\ref{fig:r_ph}, we plot the radius of AS photon ring as a function of parameter $\xi$ in comparison with Schwarzschild.  

\subsection{Lensing}  
 
Let us now investigate the deflection angle of photons due to gravitational effects in the AS spacetime. 
From equations of motion (\ref{phi_dot}) and (\ref{r_dot}), we get
\begin{equation}
    \frac{d\phi}{dr}=\pm\frac{1}{r^2}\Bigg[\frac{1}{b^2}-\frac{1}{r^2}\left(1-\frac{2M(r)}{r}\right)\Bigg]^{-1/2},
\end{equation}
where $b=L/E$ is the impact parameter. 
Integrating the equation above, we can find the deflection angle experienced by a photon with impact parameter $b$  when it passes by close to the compact object  
\be 
 \Delta\phi=2\int_{r_1}^{\infty}\frac{dr}{r^2}\Bigg[\frac{1}{b^2}-\frac{1}{r^2}\left(1-\frac{2M}{r}\right)\Bigg]^{-1/2},
\ee
where the range of integration is from infinity to the turning point $r_{1}$, due to reflection symmetry. Using condition $dr/d\phi|_{r_{1}}=0$, one can find $r_1$ from $V_{\rm eff}$ as $1/b^2$.

\begin{figure}[tt]
   \begin{center}
        \includegraphics[width=8cm]{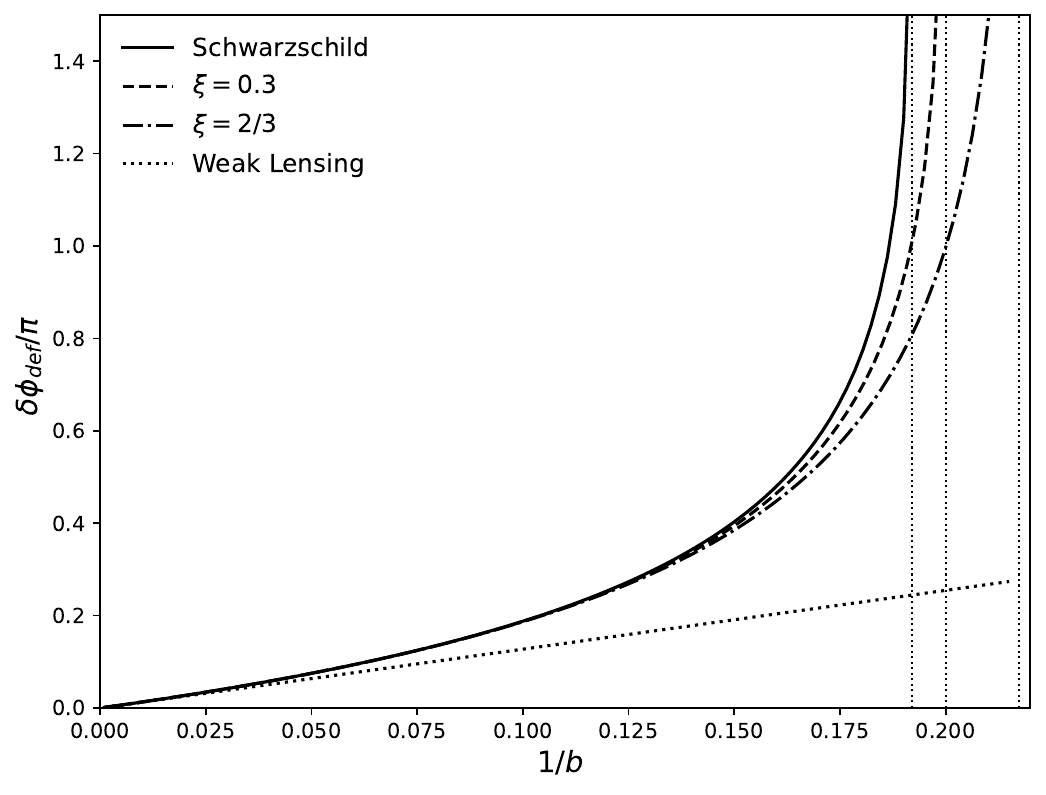}    
  \end{center}
    \caption{ The deflection angle $\delta\phi=\Delta\phi-\pi$ as a function of the inverse of the impact parameter for the ASBH as compared to Schwarzschild. The dotted line represents the weak lensing limit. The vertical dotted lines represent the critical values of impact parameters $b_{crit}$ at which photons get captured. }\label{fig:def_angle}
\end{figure}

Now we introduce a new variable $w=b/r$ and rewrite $\Delta \phi$ as
\begin{equation}
    \Delta\phi=2\int_{0}^{w_1}dw\Big[1-w^2\left(1-\frac{2M(w)}{b}w\right)\Big]^{-1/2},
\end{equation}
where $w_1=b/r_1$ and $M(w)$ 
\be
M(w) = \frac{b^3}{6 \xi w^3} \ln \left(1 + \frac{6 M_0 \xi w^3}{b^3}\right).
\ee
Fig.~\ref{fig:def_angle}  shows the dependence of the deflection angle on the inverse of the impact parameter $1/b$ for Schwarzschild and the ASBH (with $\xi/M_0^2=0.3, 2/3$).
The vertical dotted lines in Fig.~\ref{fig:def_angle}, represent the critical impact parameter $b_{crit}(\xi)$ at which photons get captured by the photon ring. The value of $b_{crit}$ determines the shadow radius of the object as it is seen by a distant observer \cite{Bambi:2013nla}.
One can calculate $b_{crit}$ by putting $r_{ph}$ into Eq.~\eqref{r_dot} and finding the solution to $E^2-V_{\rm eff}(r_{ph})=0$, namely
 \begin{equation}
     \frac{1}{b_{crit}^2}-\frac{1}{r_{ph}^2}\left(1-\frac{2M(r_{ph})}{r_{ph}}\right)=0.
 \end{equation}
In Fig.~\ref{fig:def_angle}, for the ASBH with $\xi/M_0^2=2/3$ we get $b_{crit}\approx4.600$ and for $\xi/M_0^2=0.3$ we get $b_{crit}\approx5.000$ as compared to Schwarzschild's value $b_{crit}=3\sqrt{3}\approx5.196$. These results illustrate that the shadow around the AS is smaller than Schwarzschild case \cite{Sanchez:2024sdm}. In Fig.~\ref{fig:def_angle}, we can also see that 
the deflection angles coincide in the weak field limit and
only start substantially deviating in the strong field regime close to the critical value of the impact parameter.  

\section{Motion of massive particles}\label{sec4}

For massive test particles we have the same constants of motions as in Eqs.~\eqref{t_dot} and \eqref{phi_dot} and 
now, using the normalization $g_{\mu\nu}\dot{x}^{\mu}\dot{x}^{\nu}=-1$ for timelike geodesics, we can write the equation of motion for radial coordinate as
\begin{equation}
    \dot{r}^2=E^2-\left(1+\frac{L^2}{r^2}\right)\left(1-\frac{2M(r)}{r}\right).
\end{equation}
Thus the effective potential for massive particles is
\begin{equation}\label{Veff_massive}
    V_{\rm eff}=\left(1+\frac{L^2}{r^2}\right)\left(1-\frac{2M(r)}{r}\right).
\end{equation}

\begin{figure}[tt]
   \begin{center}
        \includegraphics[width=9cm]{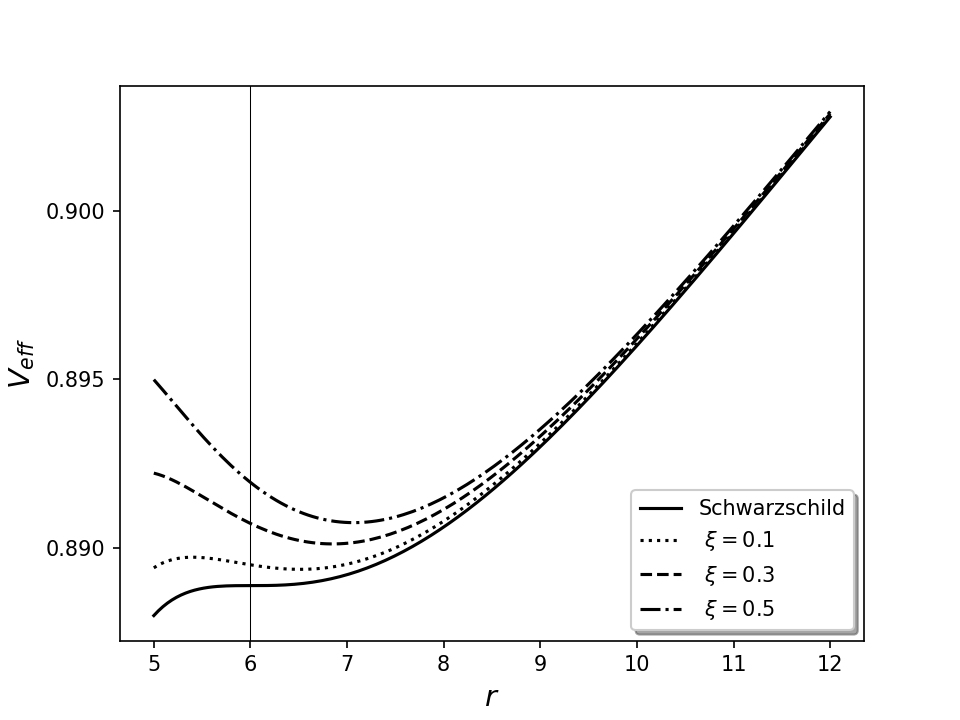}    
  \end{center}
    \caption{  The effective potential for timelike geodesics in the equatorial plane of the ASBH for three values of $\xi$ as a function of the radial coordinate in units of $M_0$. The solid line represents the effective potential for the Schwarzschild metric, and the vertical gray line shows the position of $r_{isco}$ for the Schwarzschild black hole. All plots are obtained fixing $L=12$.}\label{fig:AS_Veff_massive}
\end{figure}
In Fig.~\ref{fig:AS_Veff_massive} it is shown the effective potential for massive test particles \eqref{Veff_massive} as a function of the radial coordinate for different values of $\xi$ compared to the Schwarzschild case. Imposing the conditions for circular orbits  $V_{\rm eff}=E^2$ and $V'_{\rm eff}=0$ we find the angular momentum $L(r)$ and energy $E(r)$ for particular circular orbit as    
\begin{eqnarray}
 L^2&=&\frac{r^2(M-M' r)}{M' r+r-3M},
 \label{ang_momentum} \\
    E^2&=&\frac{(r-2M)^2}{r(M'r+r-3M)}.
    \label{energy}
\end{eqnarray}
Using the  equations of motion we obtain the angular velocity of test particles as 
\begin{equation}
    \Omega=\frac{d\phi}{dt}=\sqrt{\frac{M-M'r}{r^3}}.
    \label{omega}
\end{equation}
 
\begin{figure}[tt]
   \begin{center}
        \includegraphics[width=9cm]{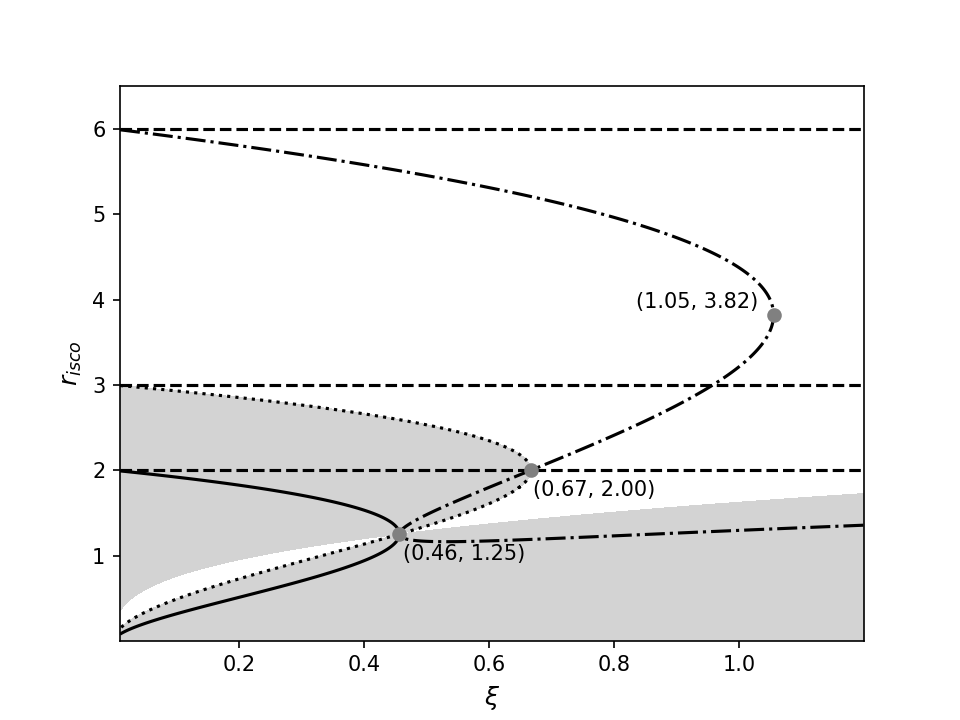}    
  \end{center}
    \caption{ The dash-dotted line represents the AS innermost stable circular orbit radius as a function of $\xi$ in units of $M_0$, the upper horizontal dashed line is $r_{isco}$ in the Schwarzschild case. The gray region corresponds to the $L^2<0$ where orbits do not exist. Also, the horizons and photon sphere for the ASBH and Schwarzschild are shown by solid, dotted, and dashed lines. The gray dot at $\xi\simeq 1.05$ highlights the limiting case for which the two marginally bound orbits of the ASBH coincide. }\label{fig:r_isco}
\end{figure}

Stable circular orbits can extend inwards only until a limiting radius called the innermost radius for circular orbits (isco). To determine the value of the radius of the isco $r_{isco}$, the condition for marginal stability of circular orbits $V_{\rm eff}''=0$ must be imposed. 
Using the explicit form of $M(r)$ for the ASBH given in Eq.~\eqref{mass_function}  we get
\begin{equation}
 M'' =  \frac{6M}{r^2} - \frac{6M_0r(r^3+15M_0\xi)}{(r^3+6M_0\xi)^2},
\end{equation}
the above equation simplifies to
\bea
    0&=& 9M_0r^3(r^3 + 12M_0\xi - 4M_0r^2)+ \nonumber \\
    && - M(r) \Big[ 
        8(r^6 + 36M_0^2\xi^2 + 12r^3M_0\xi)+ \nonumber \\
    &&  - 6M_0r^2(5r^3 + 12M_0\xi) \Big] .
\eea
In Fig.~\ref{fig:r_isco}, we 
plot the radius of the isco for massive particles in the AS metric as a function of $\xi/M_0^2$.  We also apply the condition for the physical existence of circular orbits, $L^2>0$ to exclude any unphysical parameter space where circular orbits cannot exist. We obtain two boundaries, the first for $M'r+r-3M=0$, which corresponds to the photon sphere, 
and the second for $M-M'r=0$ due to the repulsive effects of AS. This can be understood from the fact that the pseudo force, which is proportional to $g_{tt,r}$ \cite{Berry:2020ntz},
$g_{tt,r}=-2(M-M'r)/r^2$
changes sign  at $M-M'r=0$ becoming repulsive. 

For small values of $\xi/M_0^2$, the radius of the innermost stable circular orbit tends to that of the Schwarzschild metric $r_{isco}(\xi\rightarrow0)=6M_0$, as can be seen from Fig.~\ref{fig:r_isco}. The region between the photon sphere and $r_{isco}$ describes the position of unstable circular orbits. The domain outside the $r_{isco}$ represents stable circular orbits. Interestingly in the interval $0.67\lesssim \xi/M_0^2\lesssim1.05$, the ASBH metric does not possess a horizon and photon sphere. However, we have two marginally stable circular orbits for any fixed value of $\xi/M_0^2$ in that range. 
This means that particles in the accretion disk that have been plunging towards the central object after passing the isco can circularize again at smaller radii as a consequence of the repulsive effects due to weakening of gravity in Asymptotic Safety. 
Interestingly, for $\xi\gtrsim 1.05 M_0^2$, the repulsive effects of AS are such that  
marginally stable circular orbits do not occur and we have stable circular orbits all the way from infinity, much like the Newtonian case.

\subsection{Radiative Flux and Spectral Luminosity}
We now focus on the electromagnetic signatures of the accretion disk around the ASBH metric such as the radiative flux, efficiency, spectral, and differential luminosity. These signatures of the accretion disk can be used to test the possibility of distinguishing various models alternative to Schwarzschild and Kerr
\cite{Boshkayev_2020,Boshkayev_2022,Boshkayev:2023fft,D_Agostino_2023,PhysRevD.104.084009,Bambhaniya_2022}. In the following 
we make use of the formalism initially developed by Novikov \& Thorne \cite{Novikov:1973kta} and Page \& Thorne \cite{1974ApJ...191..499P}. 
The radiative flux of the accretion disk is given by
\begin{equation}
    \mathcal{F}(r)=-\frac{\dot{m}}{4\pi\sqrt{-g}}\frac{\Omega'}{(E-\Omega L)^2}\int_{r_{isco}}^r(E-\Omega L)L'({\Tilde{r}}) d\Tilde{r},
\end{equation} 
where $\dot{m}$ is the mass accretion rate, which is assumed to be constant, $g$ is the determinant of the three-sub-space metric $(t,r,\phi)$ 
and $L (r)$, $E (r)$ and $\Omega (r)$ 
for a circular geodesic are given by equations \eqref{ang_momentum},\eqref{energy} and \eqref{omega} respectively.

\begin{figure}[ttt]
   \begin{center}
 \includegraphics[width=9cm]{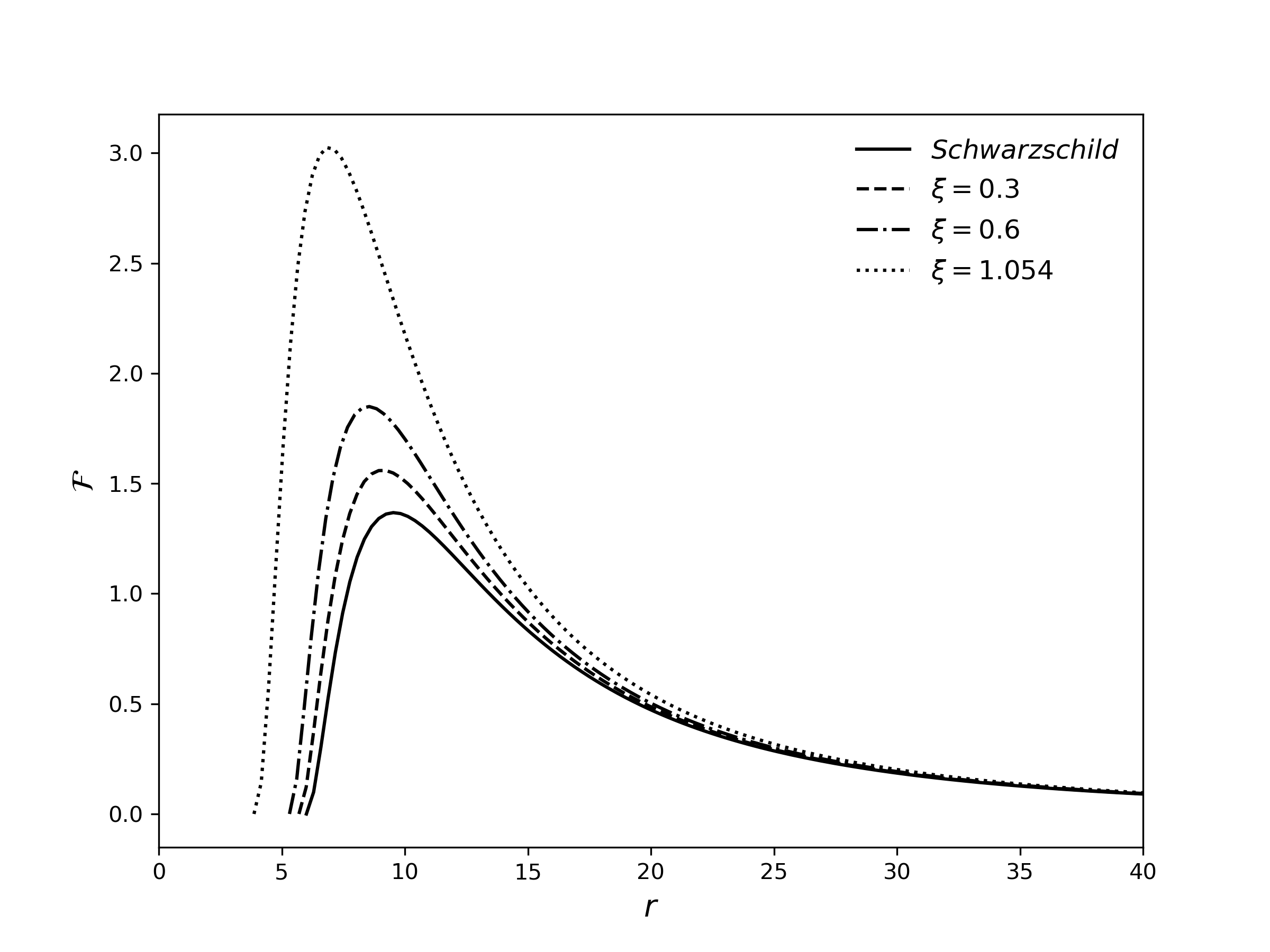}    
  \end{center}
    \caption{ Radiative Flux $\mathcal{F}$ of the thin accretion disk multiplied by $10^5$ as a function of radial coordinate r in units of $M_0$. The solid line represents the Schwarzschild black hole while the other lines corresponds to the ASBH for different values of the parameter $\xi$.  }\label{fig:Flux_As}
\end{figure}
The radiative flux $\mathcal{F}$ is not a directly observable quantity. However, it can be used to derive more useful observables such as the differential luminosity received by an observer at infinity $\mathcal{L}_{\infty}$ 
which is defined as \cite{Misner:1973prb,Novikov:1973kta}
\begin{equation}
    \frac{d\mathcal{L_{\infty}}}{d\ln{r}}=4\pi r \sqrt{g} E\mathcal{F}.
\end{equation}

Another important observable is the spectral luminosity as a function of radiation frequency. Approximating the emission of light by the gas in the accretion disk to a black body spectrum, it follows \cite{Boshkayev_2020}

\begin{equation}
    \nu \mathcal{L}_{\nu,\infty} = \frac{60}{\pi^3} 
    \int_{r_{isco}}^\infty 
    \sqrt{-g}E 
    \frac{(u^t y)^4}{\exp\left(u^t y / (\mathcal{F})^{1/4}\right) - 1} 
    \, dr,
\end{equation}
where  $y=h\nu/kT_{*}$ is a dimensionless variable, $\nu$ is the frequency of the radiation, $h$ is the Planck constant, $k$ is the Boltzmann constant, $T_{*}$ is a characteristic temperature and $u^t$ is the time component of the four-velocity vector defined as
\begin{equation}
    u^t(r)=\dot{t}=\sqrt{\frac{r}{M'r+r-3M}}.
\end{equation}
In Fig.~\ref{fig:Flux_As}, we numerically calculate radiative flux in dependence of radial distance. The figure shows that for higher values of $\xi/M_0^2$ the radiative flux emitted at small radii is larger than the corresponding flux for Schwarzschild.
Since $r_{isco}$ in the ASBH is smaller than Schwarzschild 
test particles 
can orbit closer to the center thus emit a larger flux of radiation, before falling into the compact object. 
This in turn suggests that an accretion disk surrounding an ASBH would be more luminous than that surrounding a Schwarzschild black hole of the same mass.

\begin{figure}[tt]
   \begin{center}
        \includegraphics[width=9cm]{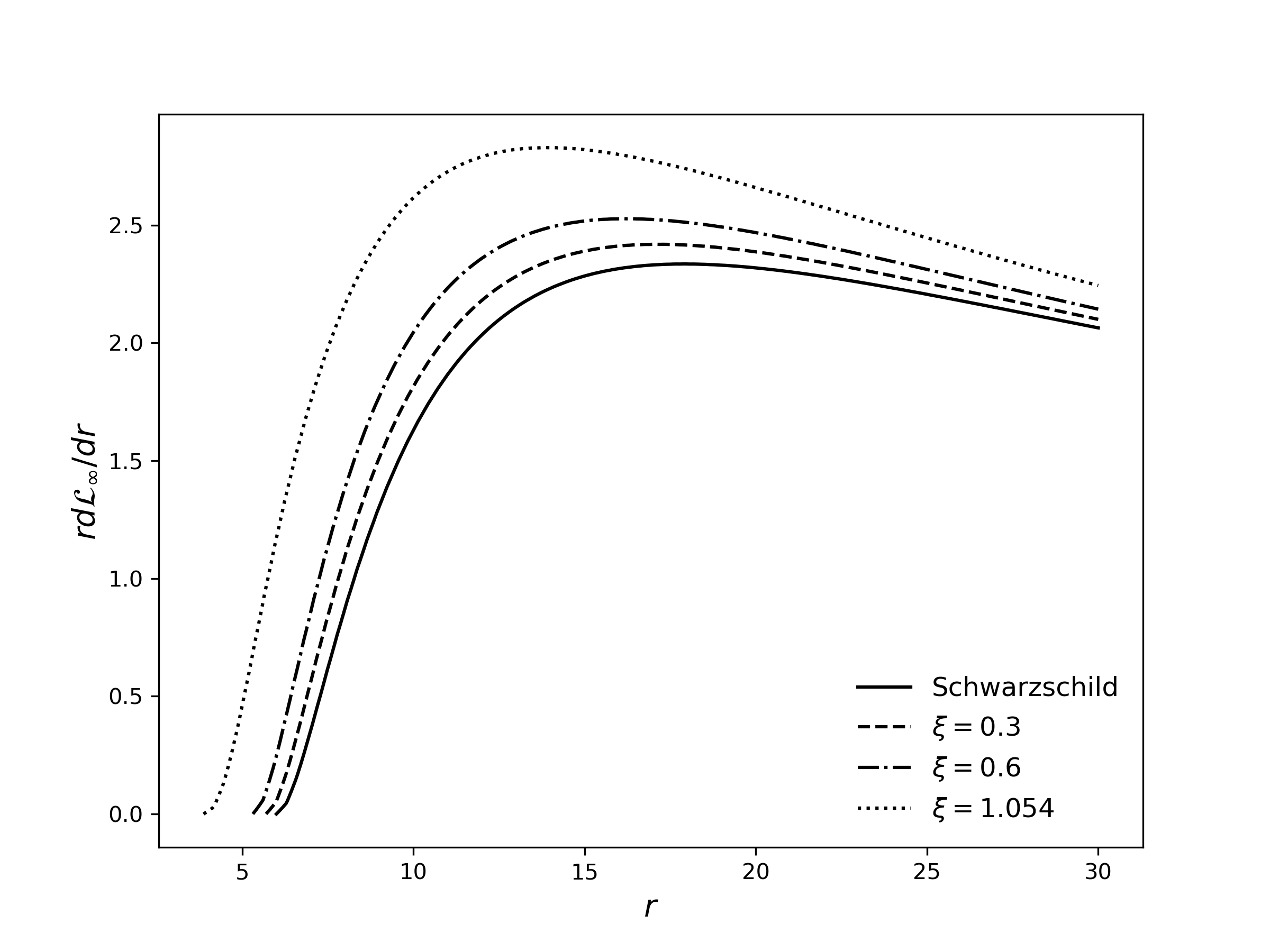}    
  \end{center}
    \caption{ Differential luminosity of the thin accretion disk multiplied by $10^2$ as a function of radial coordinate r in units of $M_0$. The solid line represents the Schwarzschild black hole while the other lines corresponds to the ASBH for different values of the parameter $\xi$.   }\label{fig:dLum}
\end{figure}

In Fig.~\ref{fig:dLum}, we plot the differential luminosity as a function of the radial distance. 
As expected we see that the differential luminosity increases with $\xi/M_0^2$.

\begin{figure}[tt]
   \begin{center}
        \includegraphics[width=9cm]{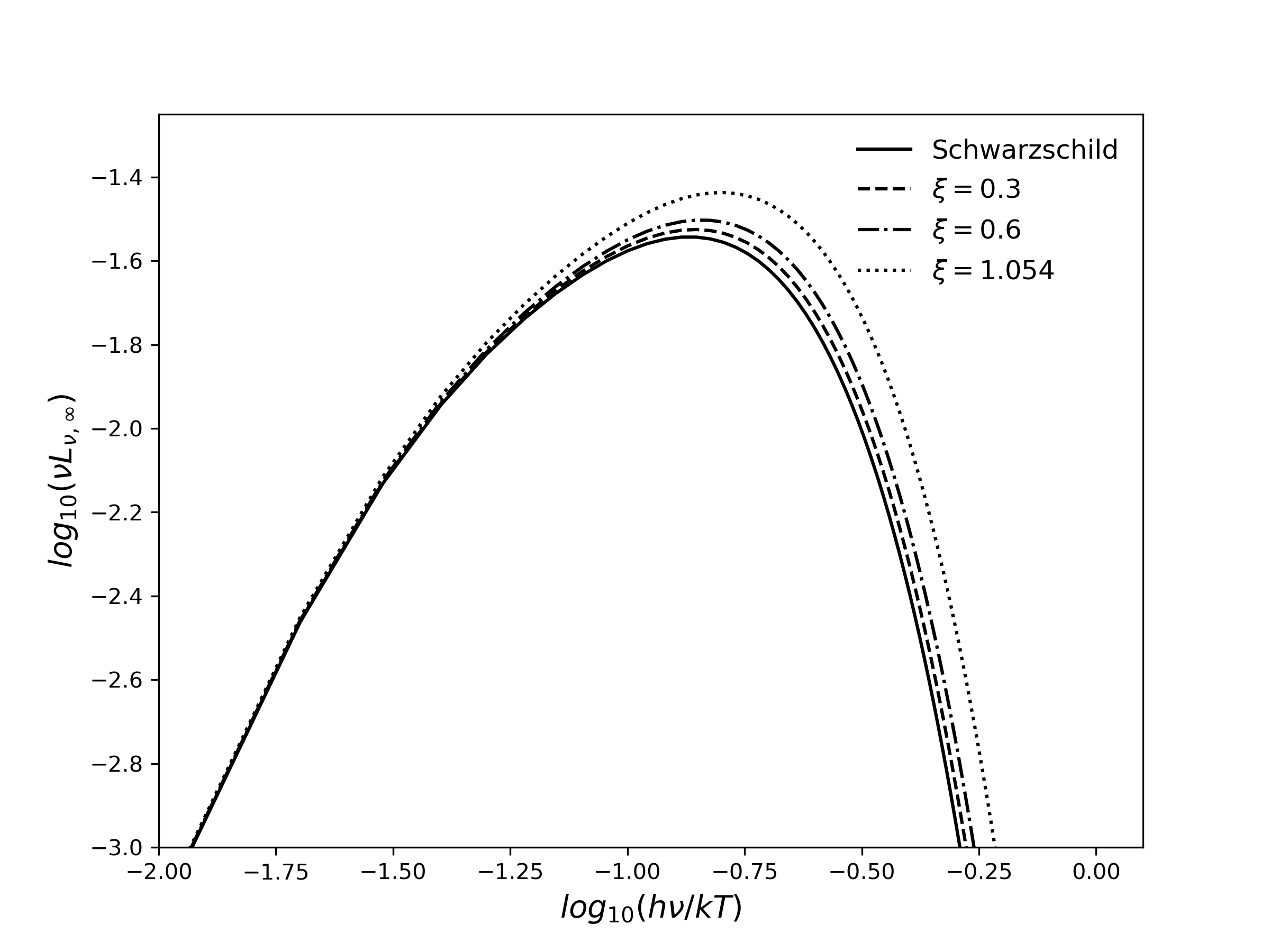}    
  \end{center}
    \caption{ Spectral Luminosity of the thin accretion disk as a function of frequency $h\nu/kT$ for $M_0=1$ in a logarithmic scale. The solid line represents the Schwarzschild black hole while the other lines corresponds to the ASBH for different values of the parameter $\xi$.     }\label{fig:Log_Lum_As}
\end{figure}

Fig.~\ref{fig:Log_Lum_As} shows another important measurable quantity: the spectral luminosity of the accretion disk as a function of radiation frequency.
We can see that the spectral luminosity from an accretion disk around the AS compact object is always larger than that of Schwarzschild. 
At higher frequencies, they are more distinguishable, while at low frequencies, they tend to coincide. 
As expected, due to the accretion disk extending closer to the center, the peak of the spectral luminosity occurs at larger frequencies, as particles from the hotter inner regions of the disk will emit more energetic photons.
Notice that we have not included the effects of particles being able to circularize again closer to the central objects for values of $\xi\in (0.67,1.05)$. This inner ring extends from the physical boundary of the collapsing object to the smaller marginally stable circular orbit.

We can also calculate the efficiency of the accretion disk in converting matter into radiation, which is given by the ratio of the energy of the photons radiated from the surface of the disk to infinity compared to the energy of the mass transported to the black hole. The efficiency of the accretion disk is proportional to the specific energy of the test particle (\ref{energy}) and can be defined as, 

\begin{equation}
    \epsilon(\xi)=1-E(r_{isco}(\xi)).
\end{equation}

\begin{figure}[tt]
   \begin{center}
        \includegraphics[width=8cm]{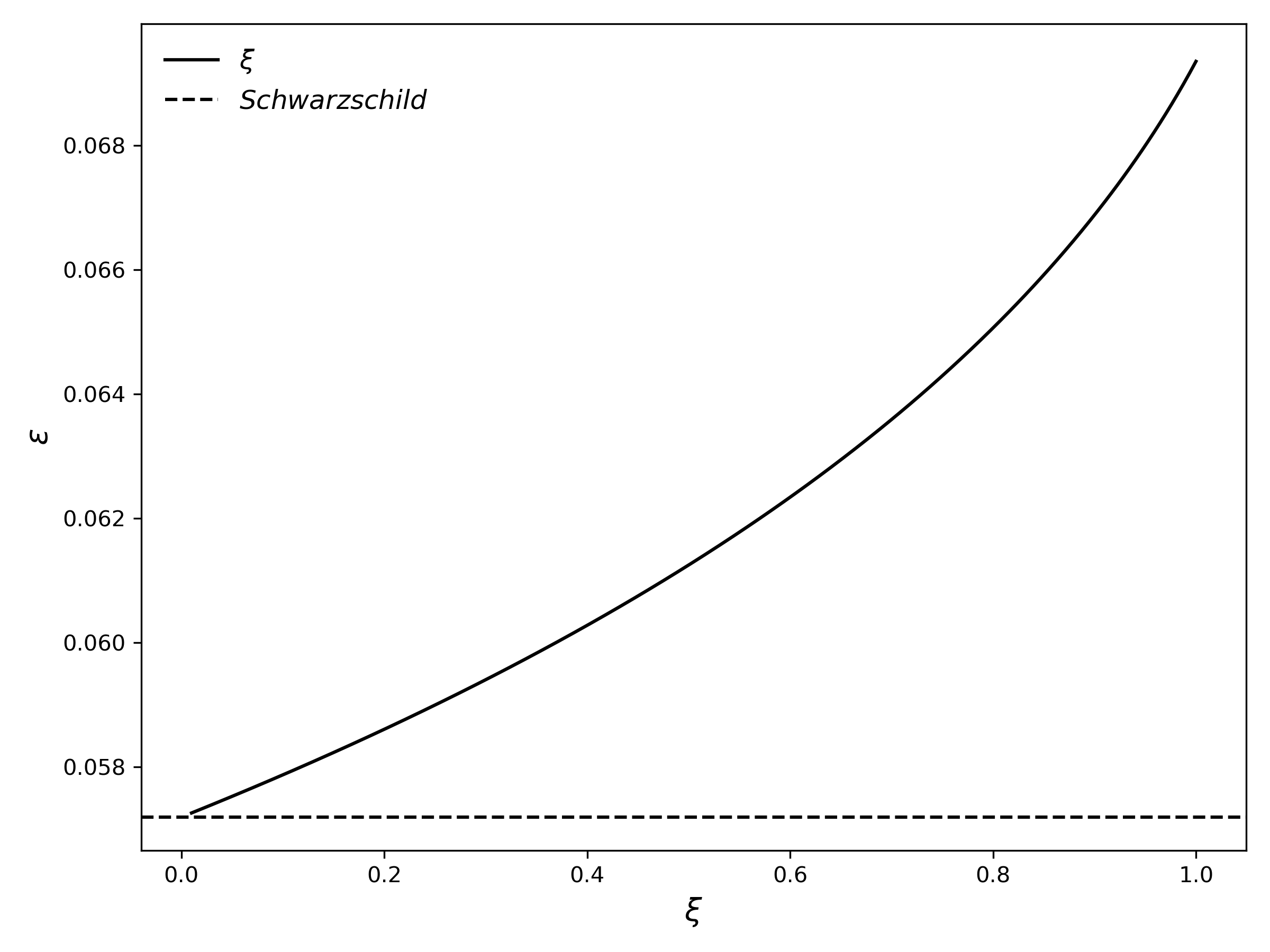}    
  \end{center}
    \caption{ The efficiency of conversion of matter into radiated energy for the ASBH accretion disk as a function of the parameter $\xi$. For $\xi\rightarrow 0$ we retrieve the Schwarzschild case (dashed horizontal line).}\label{fig:efficiency}
\end{figure}

In Fig.~\ref{fig:efficiency}, we numerically evaluate the efficiency of the accretion disk as a function of the parameter $\xi$. As $\xi \rightarrow 0$, the accretion efficiency approaches that of the Schwarzschild case $\epsilon_{Sch}\approx0.057$. However, as $\xi$ increases, the efficiency shows a noticeable growth. 

\section{K$\alpha$ Iron Line}\label{sec5}

In this section, we investigate the behavior of the K$\alpha$ iron line emitted from accretion disks in the ASBH spacetime. 
The K$\alpha$ iron line is widely used as a simple numerical tool to investigate the behavior of the spacetime geometry in the vicinity of astrophysical black hole candidates and compact objects \cite{Reynolds:2013qqa,Reynolds:2019uxi}. It is also used to test black holes in modified theories and exotic compact objects as potential alternatives to black holes \cite{Johannsen:2012ng,Bambi:2013hza,Bambi:2011jq,Bambi:2013jda,Vincent:2013uea,Tripathi:2018lhx,Bambi:2020jpe,Yang:2018wye,Zhang:2018xzj,Nampalliwar:2018iru,Zhou:2017glv,Ni:2016rhz,Cao:2016zbh}. 
In simple terms the model describes the following scenario: the hot corona located above the accretion disk illuminates the disk. The X-ray irradiation produces a reflection spectrum from the disk with imprinted fluorescent lines. The most prominent line in the X-ray spectrum is the K$\alpha$ iron line at $6.4$ KeV. Initially, the line has a narrow frequency range; however, the line measured by distant observers is broadened and skewed. This distortion is caused by relativistic effects occurring in the inner region of the disk \cite{Fabian:2000nu}. Near the center of symmetry of the geometry, the gravitational potential is relatively strong resulting in the gravitational redshift effect on the emitted photons. Additionally, in this region, the velocity of the circulating matter is extremely high, which leads to the Doppler blue- and red-shift depending on the approaching or receding sides of the disk with respect to the observer. The K$\alpha$ line was first predicted by Fabian et.al. in 1989 \cite{1989MNRAS.238..729F} and then observed in the center of the Seyfert 1 galaxy MCG-6-30-15 in 1995 \cite{Tanaka1995GravitationallyRE}.

Therefore the shape of the K$\alpha$ iron line depends on several factors including the background metric, the geometry of the emitting region, the disk's emissivity function, and the disk's inclination angle relative to a distant observer. In particular, in our case the ASBH geometry determines the trajectory of the photons emitted from the disk towards the observer, as well as the effects of gravitational redshift, the velocity of the accreting material, and the position of the innermost circular orbit, $r_{isco}$. 
We assume that the inner edge of the accretion disk is equal to $r_{isco}$ and the disk extends to a sufficiently large outer radius $r_{out}=r_{isco}+100$. However, it is important to remember that we treat the gas particles in the disk as test particles moving on geodesics while the actual geometry of the disk is unknown. Also additional emission could occur when the gas plunges in the region inside $r_{isco}$ \cite{Narayan:2005ie}. In addition it is assumed that the disk’s emissivity as a function of radius follows a power-law behavior,  $\epsilon(r)\propto r^{-\alpha}$,  where $\alpha$ is the emissivity index. The emissivity function might vary based on the model of the corona \cite{Dauser:2013xv,Chainakun:2022gcz}, however, $\alpha\sim 3$ is expected to work regardless of the specific form of the geometry \cite{Wilkins_2012}. 
By applying all the assumptions above in the context of the ASBH metric to the modeling of the K$\alpha$ iron line, we are left with two free parameters that affect the line profile: the inclination angle ($i$) and the ASBH parameter ($\xi$).

In the following, we use \texttt{Gradus.jl}\cite{gradus} written in \texttt{julia} to calculate the K$\alpha$ iron line  
The code uses ray tracing of null geodesics to compute various observational properties 
for generic spacetimes. 
We adapted the code to the ASBH geometry in order to obtain the iron line profiles for different values of the parameters $\{\xi, i\}$\footnote{The K$\alpha$ iron line profiles were calculated using the Cunningham transfer function, see \cite{1975ApJ...202..788C} for details.}.

The photon flux measured by a distant observer is denoted as 
\begin{align}
    N_{E_{obs}} &= \frac{1}{E_{obs}} \int I_{obs}(E_{obs}) d\Omega_{obs} = \notag \\
                &= \frac{1}{E_{obs}} \int g^3 I_e(E_e) d\Omega_{obs},
\end{align}
where $I_{obs}$ and $E_{obs}$ are the specific intensity of radiation and the photon energy received by the observer respectively. $I_e$ and $E_e$ are the local specific intensity and photon energy in the rest frame of the emitting gas, $d\Omega_{obs}$ is the solid angle from which a distant observer receives radiation emitted by a small element of the disk, and $g = E_{obs}/E_{e}$ is the redshift factor. According to the Liouville theorem, $I_{obs} = g^3 I_e$. It is also assumed that the radiation emitted by the disk is monochromatic, with the rest frame energy fixed at $E_{K\alpha} = 6.4$ keV, and isotropic with a power-law radial profile given by 
\begin{equation}
    I_e\propto\delta(E_e-E_{K\alpha})/r^{\alpha},
\end{equation}
where $\alpha = 3$ is the emissivity index. 
The relativistic and Doppler effects are incorporated into the calculation of $g$, while the light-bending effect is accounted for during the integration \cite{Bambi:2012tg}.

\begin{figure}[t]
    \centering
    \includegraphics[width=0.49\textwidth]{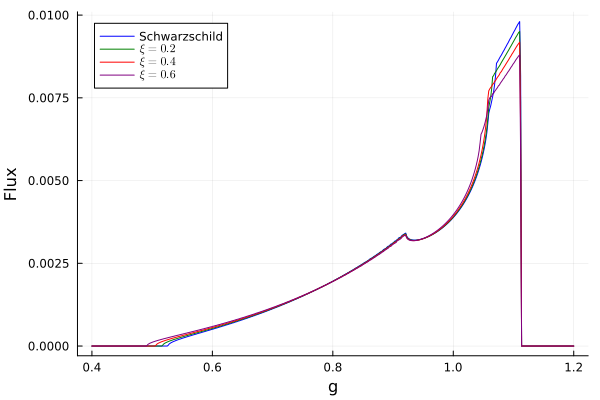}
    \hspace{0.5cm}
    \includegraphics[width=0.49\textwidth]{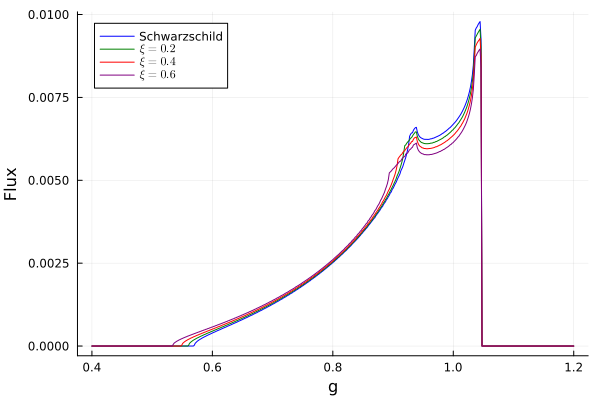}
    
    \caption{Simulation of the K$\alpha$ iron line profile in the ASBH geometry for various values of the parameter $\xi$ compared to the Schwarzschild case (solid line). The inclination angles are $45^\circ$ (top panel) and $30^\circ$ (bottom panel), and the index of the intensity profile is $\alpha = 3$.}
    \label{fig:iron_lines}
\end{figure}

\begin{figure}[t]
    \centering
    \includegraphics[width=0.49\textwidth]{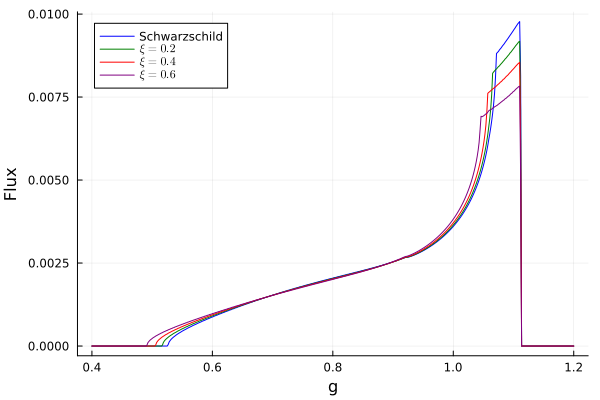}
    \hspace{0.5cm}
    \includegraphics[width=0.49\textwidth]{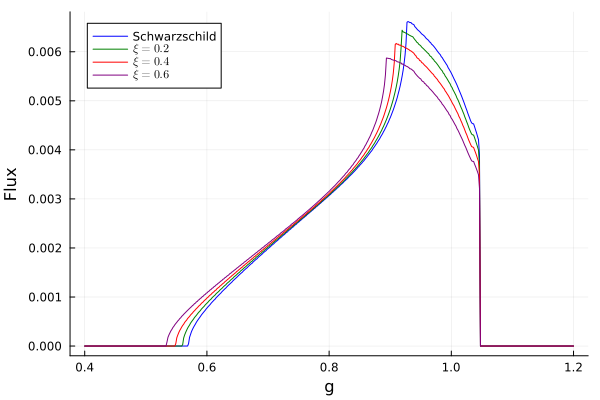}
    
    \caption{ Same as Fig.~\ref{fig:iron_lines} with the emissivity index of the intensity profile, $\alpha = 4$. Notice that the secondary peak is not visible in these plots. This is due to the fact that for $\alpha=4$ we again have large contributions from the outer portions of the disk. In fact if we consider a smaller outer edge, such as for example $r_{out}=r_{isco}+10$, the secondary peak is visible (see for example \cite{Pariev:1998ei}).}
    \label{fig:iron_lines2}
\end{figure}

In Fig.~\ref{fig:iron_lines} and \ref{fig:iron_lines2}, we plot the simulated K$\alpha$ iron line profiles in the ASBH background for various values of $\xi$, two possible inclination angles ($i=45^\circ$ and $30^\circ$) and two emissivity index ($\alpha = 3$ and $4$) and compare them to the Schwarzschild case.  
In the plots we set the outer edge of the disk at $r_{out}=r_{isco}+100$ (in units of $M_0$). Considering larger values of the outer edge includes more contributions at lower frequencies that are negligible for the AS modifications.
The profiles in the ASBH geometry closely resemble those in Schwarzschild, showing only minor differences.  
However, as the parameter $\xi$ increases, we observe that the flux of the blue-shifted peak decreases 
and the lower energy tail becomes slightly longer.  
This effect relates to the fact that in ASBH metric, as $\xi$ increases, the radius of the innermost stable circular orbit ($r_{isco}$) decreases. At smaller radii, the emitted photons experience greater gravitational redshift. In fact, a similar effect can also be observed in Kerr space-time, albeit on a larger scale for extremely fast-rotating Kerr black holes \cite{Bambi:2013hza}. For a power-law with $\alpha=3$, the lines are strongly dominated by the inner part of the disk, while on the other hand, for $\alpha = 2$, the lines are dominated by the emission from the outer part of the disk, hence asymptotic safety corrections in the $\alpha  = 2$ case would be negligible \cite{Laor1991}. Effects similar to the case of $\alpha = 3$ can also be seen for $\alpha = 4$ \cite{Pariev:1998ei}.
Notice also that for $\alpha=4$ the line profile changes significantly with respect to $\alpha=3$. However, the best fit for astrophysical objects is generally obtained for $\alpha=3$.
For $\alpha=3$, it is noteworthy that at an inclination angle of $45^\circ$, the redshifted peaks align for all profiles. However, at $30^\circ$, the redshifted peaks are distinctly different. As we increase the inclination angle, the effects of Doppler shifting and relativistic beaming become stronger, resulting in more significant blue and redshift. Note that, our calculations for the flux of the $K\alpha$ iron line is for a static black hole spacetime. In realistic scenarios, such as for a spinning counterpart of the regular black hole in asymptotic safety, the line profile can be expected to show a rich structure on the blue-shifted part \cite{Chen:2024hpw,Bromley:1996wb,Wu:2007bq}. In addition, we can also expect to observe a degeneracy between the spin of the black hole and the parameter $\xi$ since the ISCO and as a result the inner edge of the disk depends on both of these parameters.

\section{Discussion}\label{sec6}

It is generally assumed that astrophysical black hole candidates can be well described in classical GR via the standard Schwarzschild or Kerr metrics. However experimental confirmation is impossible even in principle due to the fact event horizons are not observable. Therefore one can only hope to put constraints on the validity of alternative models from observations of the environment surrounding the black hole candidates.

Recent advances in observational astrophysics have sparked significant interest in the possibility of constraining the validity of solutions in alternative theories of gravity, which can mimic the observational properties of black holes. 
Regular black holes belong to such a category of hypothetical objects. However they pose other kinds of problems as they 
generically posses Cauchy horizons, such as the inner horizon in the context of the ASBH. The existence of Cauchy horizons leads to other theoretical problems, such as mass inflation and instabilities\cite{Ori:1991zz,Poisson:1989zz,DeFelice:2024seu}. 
At the same time some authors have argued that singularities are not a limitation of GR as long as they cannot be observed \cite{Penrose:1969pc,Nakao2024}.

We investigated the motion of test particles in the vicinity of a regular black hole solution obtained within the framework of Asymptotic Safety. We examined 
circular orbits for massive particles
and photon rings depending on the free parameter $\xi$. We studied the effect of AS correction on gravitational lensing by analyzing the deflection angle of light. We investigated the electromagnetic properties of the thin accretion disk in the ASBH spacetime by calculating its radiating flux together with its differential and spectral luminosities. Additionally, we modeled the K$\alpha$ iron line profiles for various values of $\xi$ as measured by a distant observer. 

Our results indicate that the observational properties of the ASBH deviate more significantly from Schwarzschild as the value of the parameter $\xi$ increases.  
For example, the radii of the outer horizon, the photon ring, and $r_{isco}$ are smaller than those of Schwarzschild black holes and decrease as $\xi$ increases. We also see that the critical impact parameter $b_{crit}$ decreases, indicating that the photons are captured at a smaller radius. Consequently, the shadow of the AS will be smaller than Schwarzschild's in the relevant range of $\xi$.  
Similarly the properties of electromagnetic radiation emitted by the accretion disk become increasingly distinguishable from Schwarzschild as
the parameter $\xi$ grows. This is because as the value of $\xi$ rises, the size of the accretion disks increases with stable circular orbits allowed at smaller and smaller radii. 
In these inner regions, particles become more energetic and exhibit higher velocities, resulting in increased energy production and flux. Due to the stronger gravitational effects experienced by particles in the inner regions, we observe that the iron line profiles become more distorted, and low-energy tails increase as $\xi$ rises. However, for an UV cutoff at the energy scales of quantum-gravity we can expect $\xi$ to be very small.

It is important to remark that the possibility of constraining the value of $\xi$ from observations depends on the ability to accurately measure the object's $M_0$ via some independent method. In fact uncertainties on measurements of $M_0$ would impact the accuracy of the constraints on $\xi$. Furthermore our analysis is limited to the non rotating case, while for the study of astrophysical compact objects rotation can not be ignored. Indeed the existence of another free parameter, i.e. the object's angular momentum $J$, with its own experimental constraints (see e.g. \cite{Brenneman:2006hw,Tiwari:2018qch}), would reduce the ability to test the validity of the ASBH model. In addition we must mention that there exist the possibility of a degeneracy in the measurement of $J$ and $\xi$, i.e. a situation where different values of the pair $\{J,\xi\}$ produce the same measurable effects for a given observation. In this case one would need two separate experiments to provide independent values for $J$ and $\xi$ for a given compact object in order to break such degeneracy \cite{Li:2013jra,Bambi:2012ku,Bambi:2012zg,Abdikamalov:2019ztb}.

It is expected that improvements in the precision of astrophysical observations in the future will allow us to distinguish them and constrain the free parameters. As an outlook, it would be interesting to study K$\alpha$ iron line profiles of rotating regular black hole in asymptotic safety, test them against Kerr black holes, and use astrophysical data to constrain free parameter.

\section*{Acknowledgment}

The authors acknowledge support from Nazarbayev University Faculty Development Competitive Research Grant No. 040225FD4737, `{\em Modifications of General Relativity in the strong curvature regime and their implications for black holes and cosmology}'.

\bibliography{ref}

\begin{thebibliography}{94}
\expandafter\ifx\csname natexlab\endcsname\relax\def\natexlab#1{#1}\fi
\expandafter\ifx\csname bibnamefont\endcsname\relax
  \def\bibnamefont#1{#1}\fi
\expandafter\ifx\csname bibfnamefont\endcsname\relax
  \def\bibfnamefont#1{#1}\fi
\expandafter\ifx\csname citenamefont\endcsname\relax
  \def\citenamefont#1{#1}\fi
\expandafter\ifx\csname url\endcsname\relax
  \def\url#1{\texttt{#1}}\fi
\expandafter\ifx\csname urlprefix\endcsname\relax\def\urlprefix{URL }\fi
\providecommand{\bibinfo}[2]{#2}
\providecommand{\eprint}[2][]{\url{#2}}

\bibitem[{\citenamefont{Abbott et~al.}(2016)}]{PhysRevLett.116.061102}
\bibinfo{author}{\bibfnamefont{B.~P.} \bibnamefont{Abbott}} \bibnamefont{et~al.} (\bibinfo{collaboration}{LIGO Scientific, Virgo}), \bibinfo{journal}{Phys. Rev. Lett.} \textbf{\bibinfo{volume}{116}}, \bibinfo{pages}{061102} (\bibinfo{year}{2016}), \eprint{1602.03837}.

\bibitem[{\citenamefont{Akiyama et~al.}(2019{\natexlab{a}})}]{EventHorizonTelescope:2019dse}
\bibinfo{author}{\bibfnamefont{K.}~\bibnamefont{Akiyama}} \bibnamefont{et~al.} (\bibinfo{collaboration}{Event Horizon Telescope}), \bibinfo{journal}{Astrophys. J. Lett.} \textbf{\bibinfo{volume}{875}}, \bibinfo{pages}{L1} (\bibinfo{year}{2019}{\natexlab{a}}), \eprint{1906.11238}.

\bibitem[{\citenamefont{Akiyama et~al.}(2019{\natexlab{b}})}]{EventHorizonTelescope:2019ths}
\bibinfo{author}{\bibfnamefont{K.}~\bibnamefont{Akiyama}} \bibnamefont{et~al.} (\bibinfo{collaboration}{Event Horizon Telescope}), \bibinfo{journal}{Astrophys. J. Lett.} \textbf{\bibinfo{volume}{875}}, \bibinfo{pages}{L4} (\bibinfo{year}{2019}{\natexlab{b}}), \eprint{1906.11241}.

\bibitem[{\citenamefont{Hawking and Penrose}(1970)}]{Hawking:1970zqf}
\bibinfo{author}{\bibfnamefont{S.~W.} \bibnamefont{Hawking}} \bibnamefont{and} \bibinfo{author}{\bibfnamefont{R.}~\bibnamefont{Penrose}}, \bibinfo{journal}{Proc. Roy. Soc. Lond. A} \textbf{\bibinfo{volume}{314}}, \bibinfo{pages}{529} (\bibinfo{year}{1970}).

\bibitem[{\citenamefont{Senovilla}(1998)}]{Senovilla:1998oua}
\bibinfo{author}{\bibfnamefont{J.~M.~M.} \bibnamefont{Senovilla}}, \bibinfo{journal}{Gen. Rel. Grav.} \textbf{\bibinfo{volume}{30}}, \bibinfo{pages}{701} (\bibinfo{year}{1998}), \eprint{1801.04912}.

\bibitem[{\citenamefont{Cardoso and Pani}(2019)}]{Cardoso:2019rvt}
\bibinfo{author}{\bibfnamefont{V.}~\bibnamefont{Cardoso}} \bibnamefont{and} \bibinfo{author}{\bibfnamefont{P.}~\bibnamefont{Pani}}, \bibinfo{journal}{Living Rev. Rel.} \textbf{\bibinfo{volume}{22}}, \bibinfo{pages}{4} (\bibinfo{year}{2019}), \eprint{1904.05363}.

\bibitem[{\citenamefont{Bambi}(2023)}]{Bambi:2023try}
\bibinfo{editor}{\bibfnamefont{C.}~\bibnamefont{Bambi}}, ed., \emph{\bibinfo{title}{{Regular Black Holes. Towards a New Paradigm of Gravitational Collapse}}}, Springer Series in Astrophysics and Cosmology (\bibinfo{publisher}{Springer}, \bibinfo{year}{2023}), ISBN \bibinfo{isbn}{978-981-99-1595-8, 978-981-99-1598-9, 978-981-99-1596-5}, \eprint{2307.13249}.

\bibitem[{\citenamefont{Lan et~al.}(2023)\citenamefont{Lan, Yang, Guo, and Miao}}]{Lan_2023}
\bibinfo{author}{\bibfnamefont{C.}~\bibnamefont{Lan}}, \bibinfo{author}{\bibfnamefont{H.}~\bibnamefont{Yang}}, \bibinfo{author}{\bibfnamefont{Y.}~\bibnamefont{Guo}}, \bibnamefont{and} \bibinfo{author}{\bibfnamefont{Y.-G.} \bibnamefont{Miao}}, \bibinfo{journal}{International Journal of Theoretical Physics} \textbf{\bibinfo{volume}{62}} (\bibinfo{year}{2023}), ISSN \bibinfo{issn}{1572-9575}, \urlprefix\url{http://dx.doi.org/10.1007/s10773-023-05454-1}.

\bibitem[{\citenamefont{Visser and Wiltshire}(2004)}]{Visser:2003ge}
\bibinfo{author}{\bibfnamefont{M.}~\bibnamefont{Visser}} \bibnamefont{and} \bibinfo{author}{\bibfnamefont{D.~L.} \bibnamefont{Wiltshire}}, \bibinfo{journal}{Class. Quant. Grav.} \textbf{\bibinfo{volume}{21}}, \bibinfo{pages}{1135} (\bibinfo{year}{2004}), \eprint{gr-qc/0310107}.

\bibitem[{\citenamefont{Olivares et~al.}(2020)\citenamefont{Olivares, Younsi, Fromm, De~Laurentis, Porth, Mizuno, Falcke, Kramer, and Rezzolla}}]{Olivares:2018abq}
\bibinfo{author}{\bibfnamefont{H.}~\bibnamefont{Olivares}}, \bibinfo{author}{\bibfnamefont{Z.}~\bibnamefont{Younsi}}, \bibinfo{author}{\bibfnamefont{C.~M.} \bibnamefont{Fromm}}, \bibinfo{author}{\bibfnamefont{M.}~\bibnamefont{De~Laurentis}}, \bibinfo{author}{\bibfnamefont{O.}~\bibnamefont{Porth}}, \bibinfo{author}{\bibfnamefont{Y.}~\bibnamefont{Mizuno}}, \bibinfo{author}{\bibfnamefont{H.}~\bibnamefont{Falcke}}, \bibinfo{author}{\bibfnamefont{M.}~\bibnamefont{Kramer}}, \bibnamefont{and} \bibinfo{author}{\bibfnamefont{L.}~\bibnamefont{Rezzolla}}, \bibinfo{journal}{Mon. Not. Roy. Astron. Soc.} \textbf{\bibinfo{volume}{497}}, \bibinfo{pages}{521} (\bibinfo{year}{2020}), \eprint{1809.08682}.

\bibitem[{\citenamefont{Visser}(1989)}]{Visser:1989kh}
\bibinfo{author}{\bibfnamefont{M.}~\bibnamefont{Visser}}, \bibinfo{journal}{Phys. Rev. D} \textbf{\bibinfo{volume}{39}}, \bibinfo{pages}{3182} (\bibinfo{year}{1989}), \eprint{0809.0907}.

\bibitem[{\citenamefont{Casadio}(2021)}]{universe7120478}
\bibinfo{author}{\bibfnamefont{R.}~\bibnamefont{Casadio}}, \bibinfo{journal}{Universe} \textbf{\bibinfo{volume}{7}}, \bibinfo{pages}{478} (\bibinfo{year}{2021}).

\bibitem[{\citenamefont{Urmanov et~al.}(2024)\citenamefont{Urmanov, Chakrabarty, and Malafarina}}]{Urmanov:2024qai}
\bibinfo{author}{\bibfnamefont{A.}~\bibnamefont{Urmanov}}, \bibinfo{author}{\bibfnamefont{H.}~\bibnamefont{Chakrabarty}}, \bibnamefont{and} \bibinfo{author}{\bibfnamefont{D.}~\bibnamefont{Malafarina}}, \bibinfo{journal}{Phys. Rev. D} \textbf{\bibinfo{volume}{110}}, \bibinfo{pages}{044030} (\bibinfo{year}{2024}), \eprint{2406.04813}.

\bibitem[{\citenamefont{Bambi}(2017)}]{Bambi:2015kza}
\bibinfo{author}{\bibfnamefont{C.}~\bibnamefont{Bambi}}, \bibinfo{journal}{Rev. Mod. Phys.} \textbf{\bibinfo{volume}{89}}, \bibinfo{pages}{025001} (\bibinfo{year}{2017}), \eprint{1509.03884}.

\bibitem[{\citenamefont{Chakrabarty and Tang}(2023)}]{Chakrabarty:2022fbd}
\bibinfo{author}{\bibfnamefont{H.}~\bibnamefont{Chakrabarty}} \bibnamefont{and} \bibinfo{author}{\bibfnamefont{Y.}~\bibnamefont{Tang}}, \bibinfo{journal}{Phys. Rev. D} \textbf{\bibinfo{volume}{107}}, \bibinfo{pages}{084020} (\bibinfo{year}{2023}), \eprint{2204.06807}.

\bibitem[{\citenamefont{Joshi et~al.}(2013)\citenamefont{Joshi, Malafarina, and Narayan}}]{Joshi_2013}
\bibinfo{author}{\bibfnamefont{P.~S.} \bibnamefont{Joshi}}, \bibinfo{author}{\bibfnamefont{D.}~\bibnamefont{Malafarina}}, \bibnamefont{and} \bibinfo{author}{\bibfnamefont{R.}~\bibnamefont{Narayan}}, \bibinfo{journal}{Classical and Quantum Gravity} \textbf{\bibinfo{volume}{31}}, \bibinfo{pages}{015002} (\bibinfo{year}{2013}), ISSN \bibinfo{issn}{1361-6382}, \urlprefix\url{http://dx.doi.org/10.1088/0264-9381/31/1/015002}.

\bibitem[{\citenamefont{Bambi}(2013{\natexlab{a}})}]{Bambi:2013qj}
\bibinfo{author}{\bibfnamefont{C.}~\bibnamefont{Bambi}}, \bibinfo{journal}{Astron. Rev.} \textbf{\bibinfo{volume}{8}}, \bibinfo{pages}{4} (\bibinfo{year}{2013}{\natexlab{a}}), \eprint{1301.0361}.

\bibitem[{\citenamefont{Bambi}(2012{\natexlab{a}})}]{Bambi_2012a}
\bibinfo{author}{\bibfnamefont{C.}~\bibnamefont{Bambi}}, \bibinfo{journal}{Physical Review D} \textbf{\bibinfo{volume}{85}} (\bibinfo{year}{2012}{\natexlab{a}}), ISSN \bibinfo{issn}{1550-2368}, \urlprefix\url{http://dx.doi.org/10.1103/PhysRevD.85.043002}.

\bibitem[{\citenamefont{{Bardeen}}(1968)}]{1968qtr..conf...87B}
\bibinfo{author}{\bibfnamefont{J.}~\bibnamefont{{Bardeen}}}, in \emph{\bibinfo{booktitle}{Proceedings of the 5th International Conference on Gravitation and the Theory of Relativity}} (\bibinfo{year}{1968}), p.~\bibinfo{pages}{87}.

\bibitem[{\citenamefont{{Poisson} and {Israel}}(1988)}]{1988CQGra...5L.201P}
\bibinfo{author}{\bibfnamefont{E.}~\bibnamefont{{Poisson}}} \bibnamefont{and} \bibinfo{author}{\bibfnamefont{W.}~\bibnamefont{{Israel}}}, \bibinfo{journal}{Classical and Quantum Gravity} \textbf{\bibinfo{volume}{5}}, \bibinfo{pages}{L201} (\bibinfo{year}{1988}).

\bibitem[{\citenamefont{Dymnikova}(1992)}]{Dymnikova:1992ux}
\bibinfo{author}{\bibfnamefont{I.}~\bibnamefont{Dymnikova}}, \bibinfo{journal}{Gen. Rel. Grav.} \textbf{\bibinfo{volume}{24}}, \bibinfo{pages}{235} (\bibinfo{year}{1992}).

\bibitem[{\citenamefont{Hayward}(2006)}]{Hayward_2006}
\bibinfo{author}{\bibfnamefont{S.~A.} \bibnamefont{Hayward}}, \bibinfo{journal}{Physical Review Letters} \textbf{\bibinfo{volume}{96}} (\bibinfo{year}{2006}), ISSN \bibinfo{issn}{1079-7114}, \urlprefix\url{http://dx.doi.org/10.1103/PhysRevLett.96.031103}.

\bibitem[{\citenamefont{Ayon-Beato and Garcia}(1998)}]{Ayon-Beato:1998hmi}
\bibinfo{author}{\bibfnamefont{E.}~\bibnamefont{Ayon-Beato}} \bibnamefont{and} \bibinfo{author}{\bibfnamefont{A.}~\bibnamefont{Garcia}}, \bibinfo{journal}{Phys. Rev. Lett.} \textbf{\bibinfo{volume}{80}}, \bibinfo{pages}{5056} (\bibinfo{year}{1998}), \eprint{gr-qc/9911046}.

\bibitem[{\citenamefont{Bronnikov}(2001)}]{Bronnikov:2000vy}
\bibinfo{author}{\bibfnamefont{K.~A.} \bibnamefont{Bronnikov}}, \bibinfo{journal}{Phys. Rev. D} \textbf{\bibinfo{volume}{63}}, \bibinfo{pages}{044005} (\bibinfo{year}{2001}), \eprint{gr-qc/0006014}.

\bibitem[{\citenamefont{Bambi et~al.}(2017)\citenamefont{Bambi, Modesto, and Rachwa\l{}}}]{Bambi:2016wdn}
\bibinfo{author}{\bibfnamefont{C.}~\bibnamefont{Bambi}}, \bibinfo{author}{\bibfnamefont{L.}~\bibnamefont{Modesto}}, \bibnamefont{and} \bibinfo{author}{\bibfnamefont{L.}~\bibnamefont{Rachwa\l{}}}, \bibinfo{journal}{JCAP} \textbf{\bibinfo{volume}{05}}, \bibinfo{pages}{003} (\bibinfo{year}{2017}), \eprint{1611.00865}.

\bibitem[{\citenamefont{Chakrabarty et~al.}(2018)\citenamefont{Chakrabarty, Benavides-Gallego, Bambi, and Modesto}}]{Chakrabarty:2017ysw}
\bibinfo{author}{\bibfnamefont{H.}~\bibnamefont{Chakrabarty}}, \bibinfo{author}{\bibfnamefont{C.~A.} \bibnamefont{Benavides-Gallego}}, \bibinfo{author}{\bibfnamefont{C.}~\bibnamefont{Bambi}}, \bibnamefont{and} \bibinfo{author}{\bibfnamefont{L.}~\bibnamefont{Modesto}}, \bibinfo{journal}{JHEP} \textbf{\bibinfo{volume}{03}}, \bibinfo{pages}{013} (\bibinfo{year}{2018}), \eprint{1711.07198}.

\bibitem[{\citenamefont{Jusufi et~al.}(2020)\citenamefont{Jusufi, Jamil, Chakrabarty, Wu, Bambi, and Wang}}]{Jusufi:2019caq}
\bibinfo{author}{\bibfnamefont{K.}~\bibnamefont{Jusufi}}, \bibinfo{author}{\bibfnamefont{M.}~\bibnamefont{Jamil}}, \bibinfo{author}{\bibfnamefont{H.}~\bibnamefont{Chakrabarty}}, \bibinfo{author}{\bibfnamefont{Q.}~\bibnamefont{Wu}}, \bibinfo{author}{\bibfnamefont{C.}~\bibnamefont{Bambi}}, \bibnamefont{and} \bibinfo{author}{\bibfnamefont{A.}~\bibnamefont{Wang}}, \bibinfo{journal}{Phys. Rev. D} \textbf{\bibinfo{volume}{101}}, \bibinfo{pages}{044035} (\bibinfo{year}{2020}), \eprint{1911.07520}.

\bibitem[{\citenamefont{Fan and Wang}(2016)}]{Fan:2016hvf}
\bibinfo{author}{\bibfnamefont{Z.-Y.} \bibnamefont{Fan}} \bibnamefont{and} \bibinfo{author}{\bibfnamefont{X.}~\bibnamefont{Wang}}, \bibinfo{journal}{Phys. Rev. D} \textbf{\bibinfo{volume}{94}}, \bibinfo{pages}{124027} (\bibinfo{year}{2016}), \eprint{1610.02636}.

\bibitem[{\citenamefont{Bronnikov}(2017)}]{Bronnikov:2017tnz}
\bibinfo{author}{\bibfnamefont{K.~A.} \bibnamefont{Bronnikov}}, \bibinfo{journal}{Phys. Rev. D} \textbf{\bibinfo{volume}{96}}, \bibinfo{pages}{128501} (\bibinfo{year}{2017}), \eprint{1712.04342}.

\bibitem[{\citenamefont{Toshmatov et~al.}(2018)\citenamefont{Toshmatov, Stuchl\'\i{}k, and Ahmedov}}]{Toshmatov:2018cks}
\bibinfo{author}{\bibfnamefont{B.}~\bibnamefont{Toshmatov}}, \bibinfo{author}{\bibfnamefont{Z.}~\bibnamefont{Stuchl\'\i{}k}}, \bibnamefont{and} \bibinfo{author}{\bibfnamefont{B.}~\bibnamefont{Ahmedov}}, \bibinfo{journal}{Phys. Rev. D} \textbf{\bibinfo{volume}{98}}, \bibinfo{pages}{028501} (\bibinfo{year}{2018}), \eprint{1807.09502}.

\bibitem[{\citenamefont{Malafarina and Toshmatov}(2022)}]{Malafarina:2022oka}
\bibinfo{author}{\bibfnamefont{D.}~\bibnamefont{Malafarina}} \bibnamefont{and} \bibinfo{author}{\bibfnamefont{B.}~\bibnamefont{Toshmatov}}, \bibinfo{journal}{Phys. Rev. D} \textbf{\bibinfo{volume}{105}}, \bibinfo{pages}{L121502} (\bibinfo{year}{2022}), \eprint{2204.04025}.

\bibitem[{\citenamefont{Frolov and Vilkovisky}(1981)}]{Frolov:1981mz}
\bibinfo{author}{\bibfnamefont{V.~P.} \bibnamefont{Frolov}} \bibnamefont{and} \bibinfo{author}{\bibfnamefont{G.~A.} \bibnamefont{Vilkovisky}}, \bibinfo{journal}{Phys. Lett. B} \textbf{\bibinfo{volume}{106}}, \bibinfo{pages}{307} (\bibinfo{year}{1981}).

\bibitem[{\citenamefont{Bonanno and Reuter}(2000)}]{Bonanno_2000}
\bibinfo{author}{\bibfnamefont{A.}~\bibnamefont{Bonanno}} \bibnamefont{and} \bibinfo{author}{\bibfnamefont{M.}~\bibnamefont{Reuter}}, \bibinfo{journal}{Physical Review D} \textbf{\bibinfo{volume}{62}} (\bibinfo{year}{2000}), ISSN \bibinfo{issn}{1089-4918}, \urlprefix\url{http://dx.doi.org/10.1103/PhysRevD.62.043008}.

\bibitem[{\citenamefont{Bambi et~al.}(2013)\citenamefont{Bambi, Malafarina, and Modesto}}]{Bambi:2013caa}
\bibinfo{author}{\bibfnamefont{C.}~\bibnamefont{Bambi}}, \bibinfo{author}{\bibfnamefont{D.}~\bibnamefont{Malafarina}}, \bibnamefont{and} \bibinfo{author}{\bibfnamefont{L.}~\bibnamefont{Modesto}}, \bibinfo{journal}{Phys. Rev. D} \textbf{\bibinfo{volume}{88}}, \bibinfo{pages}{044009} (\bibinfo{year}{2013}), \eprint{1305.4790}.

\bibitem[{\citenamefont{Malafarina}(2017)}]{Malafarina:2017csn}
\bibinfo{author}{\bibfnamefont{D.}~\bibnamefont{Malafarina}}, \bibinfo{journal}{Universe} \textbf{\bibinfo{volume}{3}}, \bibinfo{pages}{48} (\bibinfo{year}{2017}), \eprint{1703.04138}.

\bibitem[{\citenamefont{Carballo-Rubio et~al.}(2025)}]{Carballo-Rubio:2025fnc}
\bibinfo{author}{\bibfnamefont{R.}~\bibnamefont{Carballo-Rubio}} \bibnamefont{et~al.} (\bibinfo{year}{2025}), \eprint{2501.05505}.

\bibitem[{\citenamefont{Bonanno et~al.}(2024)\citenamefont{Bonanno, Malafarina, and Panassiti}}]{PhysRevLett.132.031401}
\bibinfo{author}{\bibfnamefont{A.}~\bibnamefont{Bonanno}}, \bibinfo{author}{\bibfnamefont{D.}~\bibnamefont{Malafarina}}, \bibnamefont{and} \bibinfo{author}{\bibfnamefont{A.}~\bibnamefont{Panassiti}}, \bibinfo{journal}{Phys. Rev. Lett.} \textbf{\bibinfo{volume}{132}}, \bibinfo{pages}{031401} (\bibinfo{year}{2024}), \urlprefix\url{https://link.aps.org/doi/10.1103/PhysRevLett.132.031401}.

\bibitem[{\citenamefont{{Markov} and {Mukhanov}}(1984)}]{1984JETPL..40.1043M}
\bibinfo{author}{\bibfnamefont{M.~A.} \bibnamefont{{Markov}}} \bibnamefont{and} \bibinfo{author}{\bibfnamefont{V.~F.} \bibnamefont{{Mukhanov}}}, \bibinfo{journal}{Soviet Journal of Experimental and Theoretical Physics Letters} \textbf{\bibinfo{volume}{40}}, \bibinfo{pages}{1043} (\bibinfo{year}{1984}).

\bibitem[{\citenamefont{Stashko}(2024)}]{Stashko_2024}
\bibinfo{author}{\bibfnamefont{O.}~\bibnamefont{Stashko}}, \bibinfo{journal}{Physical Review D} \textbf{\bibinfo{volume}{110}} (\bibinfo{year}{2024}), ISSN \bibinfo{issn}{2470-0029}, \urlprefix\url{http://dx.doi.org/10.1103/PhysRevD.110.084016}.

\bibitem[{\citenamefont{S\'anchez}(2024)}]{Sanchez:2024sdm}
\bibinfo{author}{\bibfnamefont{L.~A.} \bibnamefont{S\'anchez}} (\bibinfo{year}{2024}), \eprint{2408.00226}.

\bibitem[{\citenamefont{{Oppenheimer} and {Snyder}}(1939)}]{1939PhRv...56..455O}
\bibinfo{author}{\bibfnamefont{J.~R.} \bibnamefont{{Oppenheimer}}} \bibnamefont{and} \bibinfo{author}{\bibfnamefont{H.}~\bibnamefont{{Snyder}}}, \bibinfo{journal}{Physical Review} \textbf{\bibinfo{volume}{56}}, \bibinfo{pages}{455} (\bibinfo{year}{1939}).

\bibitem[{\citenamefont{{Datt}}(1938)}]{1938ZPhy..108..314D}
\bibinfo{author}{\bibfnamefont{B.}~\bibnamefont{{Datt}}}, \bibinfo{journal}{Zeitschrift fur Physik} \textbf{\bibinfo{volume}{108}}, \bibinfo{pages}{314} (\bibinfo{year}{1938}).

\bibitem[{\citenamefont{Bambi}(2013{\natexlab{b}})}]{Bambi:2013nla}
\bibinfo{author}{\bibfnamefont{C.}~\bibnamefont{Bambi}}, \bibinfo{journal}{Phys. Rev. D} \textbf{\bibinfo{volume}{87}}, \bibinfo{pages}{107501} (\bibinfo{year}{2013}{\natexlab{b}}), \eprint{1304.5691}.

\bibitem[{\citenamefont{Berry et~al.}(2020)\citenamefont{Berry, Simpson, and Visser}}]{Berry:2020ntz}
\bibinfo{author}{\bibfnamefont{T.}~\bibnamefont{Berry}}, \bibinfo{author}{\bibfnamefont{A.}~\bibnamefont{Simpson}}, \bibnamefont{and} \bibinfo{author}{\bibfnamefont{M.}~\bibnamefont{Visser}}, \bibinfo{journal}{Universe} \textbf{\bibinfo{volume}{7}}, \bibinfo{pages}{2} (\bibinfo{year}{2020}), \eprint{2008.13308}.

\bibitem[{\citenamefont{Boshkayev et~al.}(2020)\citenamefont{Boshkayev, Idrissov, Luongo, and Malafarina}}]{Boshkayev_2020}
\bibinfo{author}{\bibfnamefont{K.}~\bibnamefont{Boshkayev}}, \bibinfo{author}{\bibfnamefont{A.}~\bibnamefont{Idrissov}}, \bibinfo{author}{\bibfnamefont{O.}~\bibnamefont{Luongo}}, \bibnamefont{and} \bibinfo{author}{\bibfnamefont{D.}~\bibnamefont{Malafarina}}, \bibinfo{journal}{Monthly Notices of the Royal Astronomical Society} \textbf{\bibinfo{volume}{496}}, \bibinfo{pages}{1115–1123} (\bibinfo{year}{2020}), ISSN \bibinfo{issn}{1365-2966}, \urlprefix\url{http://dx.doi.org/10.1093/mnras/staa1564}.

\bibitem[{\citenamefont{Boshkayev et~al.}(2022)\citenamefont{Boshkayev, Konysbayev, Kurmanov, Luongo, and Malafarina}}]{Boshkayev_2022}
\bibinfo{author}{\bibfnamefont{K.}~\bibnamefont{Boshkayev}}, \bibinfo{author}{\bibfnamefont{T.}~\bibnamefont{Konysbayev}}, \bibinfo{author}{\bibfnamefont{Y.}~\bibnamefont{Kurmanov}}, \bibinfo{author}{\bibfnamefont{O.}~\bibnamefont{Luongo}}, \bibnamefont{and} \bibinfo{author}{\bibfnamefont{D.}~\bibnamefont{Malafarina}}, \bibinfo{journal}{The Astrophysical Journal} \textbf{\bibinfo{volume}{936}}, \bibinfo{pages}{96} (\bibinfo{year}{2022}), ISSN \bibinfo{issn}{1538-4357}, \urlprefix\url{http://dx.doi.org/10.3847/1538-4357/ac8804}.

\bibitem[{\citenamefont{Boshkayev et~al.}(2024)\citenamefont{Boshkayev, Konysbayev, Kurmanov, Luongo, Muccino, Taukenova, and Urazalina}}]{Boshkayev:2023fft}
\bibinfo{author}{\bibfnamefont{K.}~\bibnamefont{Boshkayev}}, \bibinfo{author}{\bibfnamefont{T.}~\bibnamefont{Konysbayev}}, \bibinfo{author}{\bibfnamefont{Y.}~\bibnamefont{Kurmanov}}, \bibinfo{author}{\bibfnamefont{O.}~\bibnamefont{Luongo}}, \bibinfo{author}{\bibfnamefont{M.}~\bibnamefont{Muccino}}, \bibinfo{author}{\bibfnamefont{A.}~\bibnamefont{Taukenova}}, \bibnamefont{and} \bibinfo{author}{\bibfnamefont{A.}~\bibnamefont{Urazalina}}, \bibinfo{journal}{Eur. Phys. J. C} \textbf{\bibinfo{volume}{84}}, \bibinfo{pages}{230} (\bibinfo{year}{2024}), \eprint{2307.15003}.

\bibitem[{\citenamefont{D’Agostino et~al.}(2023)\citenamefont{D’Agostino, Giambò, and Luongo}}]{D_Agostino_2023}
\bibinfo{author}{\bibfnamefont{R.}~\bibnamefont{D’Agostino}}, \bibinfo{author}{\bibfnamefont{R.}~\bibnamefont{Giambò}}, \bibnamefont{and} \bibinfo{author}{\bibfnamefont{O.}~\bibnamefont{Luongo}}, \bibinfo{journal}{Physical Review D} \textbf{\bibinfo{volume}{107}} (\bibinfo{year}{2023}), ISSN \bibinfo{issn}{2470-0029}, \urlprefix\url{http://dx.doi.org/10.1103/PhysRevD.107.043032}.

\bibitem[{\citenamefont{Boshkayev et~al.}(2021)\citenamefont{Boshkayev, Konysbayev, Kurmanov, Luongo, Malafarina, and Quevedo}}]{PhysRevD.104.084009}
\bibinfo{author}{\bibfnamefont{K.}~\bibnamefont{Boshkayev}}, \bibinfo{author}{\bibfnamefont{T.}~\bibnamefont{Konysbayev}}, \bibinfo{author}{\bibfnamefont{E.}~\bibnamefont{Kurmanov}}, \bibinfo{author}{\bibfnamefont{O.}~\bibnamefont{Luongo}}, \bibinfo{author}{\bibfnamefont{D.}~\bibnamefont{Malafarina}}, \bibnamefont{and} \bibinfo{author}{\bibfnamefont{H.}~\bibnamefont{Quevedo}}, \bibinfo{journal}{Phys. Rev. D} \textbf{\bibinfo{volume}{104}}, \bibinfo{pages}{084009} (\bibinfo{year}{2021}), \urlprefix\url{https://link.aps.org/doi/10.1103/PhysRevD.104.084009}.

\bibitem[{\citenamefont{Bambhaniya et~al.}(2022)\citenamefont{Bambhaniya, Saurabh, Jusufi, and Joshi}}]{Bambhaniya_2022}
\bibinfo{author}{\bibfnamefont{P.}~\bibnamefont{Bambhaniya}}, \bibinfo{author}{\bibnamefont{Saurabh}}, \bibinfo{author}{\bibfnamefont{K.}~\bibnamefont{Jusufi}}, \bibnamefont{and} \bibinfo{author}{\bibfnamefont{P.~S.} \bibnamefont{Joshi}}, \bibinfo{journal}{Physical Review D} \textbf{\bibinfo{volume}{105}} (\bibinfo{year}{2022}), ISSN \bibinfo{issn}{2470-0029}, \urlprefix\url{http://dx.doi.org/10.1103/PhysRevD.105.023021}.

\bibitem[{\citenamefont{Novikov and Thorne}(1973)}]{Novikov:1973kta}
\bibinfo{author}{\bibfnamefont{I.~D.} \bibnamefont{Novikov}} \bibnamefont{and} \bibinfo{author}{\bibfnamefont{K.~S.} \bibnamefont{Thorne}}, in \emph{\bibinfo{booktitle}{{Les Houches Summer School of Theoretical Physics}: {Black Holes}}} (\bibinfo{year}{1973}), pp. \bibinfo{pages}{343--550}.

\bibitem[{\citenamefont{{Page} and {Thorne}}(1974)}]{1974ApJ...191..499P}
\bibinfo{author}{\bibfnamefont{D.~N.} \bibnamefont{{Page}}} \bibnamefont{and} \bibinfo{author}{\bibfnamefont{K.~S.} \bibnamefont{{Thorne}}}, \bibinfo{journal}{\apj} \textbf{\bibinfo{volume}{191}}, \bibinfo{pages}{499} (\bibinfo{year}{1974}).

\bibitem[{\citenamefont{Misner et~al.}(1973)\citenamefont{Misner, Thorne, and Wheeler}}]{Misner:1973prb}
\bibinfo{author}{\bibfnamefont{C.~W.} \bibnamefont{Misner}}, \bibinfo{author}{\bibfnamefont{K.~S.} \bibnamefont{Thorne}}, \bibnamefont{and} \bibinfo{author}{\bibfnamefont{J.~A.} \bibnamefont{Wheeler}}, \emph{\bibinfo{title}{{Gravitation}}} (\bibinfo{publisher}{W. H. Freeman}, \bibinfo{address}{San Francisco}, \bibinfo{year}{1973}), ISBN \bibinfo{isbn}{978-0-7167-0344-0, 978-0-691-17779-3}.

\bibitem[{\citenamefont{Reynolds}(2014)}]{Reynolds:2013qqa}
\bibinfo{author}{\bibfnamefont{C.~S.} \bibnamefont{Reynolds}}, \bibinfo{journal}{Space Sci. Rev.} \textbf{\bibinfo{volume}{183}}, \bibinfo{pages}{277} (\bibinfo{year}{2014}), \eprint{1302.3260}.

\bibitem[{\citenamefont{Reynolds}(2019)}]{Reynolds:2019uxi}
\bibinfo{author}{\bibfnamefont{C.~S.} \bibnamefont{Reynolds}}, \bibinfo{journal}{Nature Astron.} \textbf{\bibinfo{volume}{3}}, \bibinfo{pages}{41} (\bibinfo{year}{2019}), \eprint{1903.11704}.

\bibitem[{\citenamefont{Johannsen and Psaltis}(2013)}]{Johannsen:2012ng}
\bibinfo{author}{\bibfnamefont{T.}~\bibnamefont{Johannsen}} \bibnamefont{and} \bibinfo{author}{\bibfnamefont{D.}~\bibnamefont{Psaltis}}, \bibinfo{journal}{Astrophys. J.} \textbf{\bibinfo{volume}{773}}, \bibinfo{pages}{57} (\bibinfo{year}{2013}), \eprint{1202.6069}.

\bibitem[{\citenamefont{Bambi and Malafarina}(2013)}]{Bambi:2013hza}
\bibinfo{author}{\bibfnamefont{C.}~\bibnamefont{Bambi}} \bibnamefont{and} \bibinfo{author}{\bibfnamefont{D.}~\bibnamefont{Malafarina}}, \bibinfo{journal}{Phys. Rev. D} \textbf{\bibinfo{volume}{88}}, \bibinfo{pages}{064022} (\bibinfo{year}{2013}), \eprint{1307.2106}.

\bibitem[{\citenamefont{Bambi and Barausse}(2011)}]{Bambi:2011jq}
\bibinfo{author}{\bibfnamefont{C.}~\bibnamefont{Bambi}} \bibnamefont{and} \bibinfo{author}{\bibfnamefont{E.}~\bibnamefont{Barausse}}, \bibinfo{journal}{Astrophys. J.} \textbf{\bibinfo{volume}{731}}, \bibinfo{pages}{121} (\bibinfo{year}{2011}), \eprint{1012.2007}.

\bibitem[{\citenamefont{Bambi}(2013{\natexlab{c}})}]{Bambi:2013jda}
\bibinfo{author}{\bibfnamefont{C.}~\bibnamefont{Bambi}}, \bibinfo{journal}{Phys. Rev. D} \textbf{\bibinfo{volume}{87}}, \bibinfo{pages}{084039} (\bibinfo{year}{2013}{\natexlab{c}}), \eprint{1303.0624}.

\bibitem[{\citenamefont{Vincent}(2013)}]{Vincent:2013uea}
\bibinfo{author}{\bibfnamefont{F.~H.} \bibnamefont{Vincent}}, \bibinfo{journal}{Class. Quant. Grav.} \textbf{\bibinfo{volume}{31}}, \bibinfo{pages}{025010} (\bibinfo{year}{2013}), \eprint{1311.3251}.

\bibitem[{\citenamefont{Tripathi et~al.}(2019)\citenamefont{Tripathi, Nampalliwar, Abdikamalov, Ayzenberg, Bambi, Dauser, Garcia, and Marinucci}}]{Tripathi:2018lhx}
\bibinfo{author}{\bibfnamefont{A.}~\bibnamefont{Tripathi}}, \bibinfo{author}{\bibfnamefont{S.}~\bibnamefont{Nampalliwar}}, \bibinfo{author}{\bibfnamefont{A.~B.} \bibnamefont{Abdikamalov}}, \bibinfo{author}{\bibfnamefont{D.}~\bibnamefont{Ayzenberg}}, \bibinfo{author}{\bibfnamefont{C.}~\bibnamefont{Bambi}}, \bibinfo{author}{\bibfnamefont{T.}~\bibnamefont{Dauser}}, \bibinfo{author}{\bibfnamefont{J.~A.} \bibnamefont{Garcia}}, \bibnamefont{and} \bibinfo{author}{\bibfnamefont{A.}~\bibnamefont{Marinucci}}, \bibinfo{journal}{Astrophys. J.} \textbf{\bibinfo{volume}{875}}, \bibinfo{pages}{56} (\bibinfo{year}{2019}), \eprint{1811.08148}.

\bibitem[{\citenamefont{Bambi et~al.}(2021)}]{Bambi:2020jpe}
\bibinfo{author}{\bibfnamefont{C.}~\bibnamefont{Bambi}} \bibnamefont{et~al.}, \bibinfo{journal}{Space Sci. Rev.} \textbf{\bibinfo{volume}{217}}, \bibinfo{pages}{65} (\bibinfo{year}{2021}), \eprint{2011.04792}.

\bibitem[{\citenamefont{Yang et~al.}(2018)\citenamefont{Yang, Ayzenberg, and Bambi}}]{Yang:2018wye}
\bibinfo{author}{\bibfnamefont{J.}~\bibnamefont{Yang}}, \bibinfo{author}{\bibfnamefont{D.}~\bibnamefont{Ayzenberg}}, \bibnamefont{and} \bibinfo{author}{\bibfnamefont{C.}~\bibnamefont{Bambi}}, \bibinfo{journal}{Phys. Rev. D} \textbf{\bibinfo{volume}{98}}, \bibinfo{pages}{044024} (\bibinfo{year}{2018}), \eprint{1806.06240}.

\bibitem[{\citenamefont{Zhang et~al.}(2018)\citenamefont{Zhang, Zhou, and Bambi}}]{Zhang:2018xzj}
\bibinfo{author}{\bibfnamefont{Y.}~\bibnamefont{Zhang}}, \bibinfo{author}{\bibfnamefont{M.}~\bibnamefont{Zhou}}, \bibnamefont{and} \bibinfo{author}{\bibfnamefont{C.}~\bibnamefont{Bambi}}, \bibinfo{journal}{Eur. Phys. J. C} \textbf{\bibinfo{volume}{78}}, \bibinfo{pages}{376} (\bibinfo{year}{2018}), \eprint{1804.07955}.

\bibitem[{\citenamefont{Nampalliwar et~al.}(2018)\citenamefont{Nampalliwar, Bambi, Kokkotas, and Konoplya}}]{Nampalliwar:2018iru}
\bibinfo{author}{\bibfnamefont{S.}~\bibnamefont{Nampalliwar}}, \bibinfo{author}{\bibfnamefont{C.}~\bibnamefont{Bambi}}, \bibinfo{author}{\bibfnamefont{K.}~\bibnamefont{Kokkotas}}, \bibnamefont{and} \bibinfo{author}{\bibfnamefont{R.}~\bibnamefont{Konoplya}}, \bibinfo{journal}{Phys. Lett. B} \textbf{\bibinfo{volume}{781}}, \bibinfo{pages}{626} (\bibinfo{year}{2018}), \eprint{1803.10819}.

\bibitem[{\citenamefont{Zhou et~al.}(2017)\citenamefont{Zhou, Bambi, Herdeiro, and Radu}}]{Zhou:2017glv}
\bibinfo{author}{\bibfnamefont{M.}~\bibnamefont{Zhou}}, \bibinfo{author}{\bibfnamefont{C.}~\bibnamefont{Bambi}}, \bibinfo{author}{\bibfnamefont{C.~A.~R.} \bibnamefont{Herdeiro}}, \bibnamefont{and} \bibinfo{author}{\bibfnamefont{E.}~\bibnamefont{Radu}}, \bibinfo{journal}{Phys. Rev. D} \textbf{\bibinfo{volume}{95}}, \bibinfo{pages}{104035} (\bibinfo{year}{2017}), \eprint{1703.06836}.

\bibitem[{\citenamefont{Ni et~al.}(2016)\citenamefont{Ni, Zhou, Cardenas-Avendano, Bambi, Herdeiro, and Radu}}]{Ni:2016rhz}
\bibinfo{author}{\bibfnamefont{Y.}~\bibnamefont{Ni}}, \bibinfo{author}{\bibfnamefont{M.}~\bibnamefont{Zhou}}, \bibinfo{author}{\bibfnamefont{A.}~\bibnamefont{Cardenas-Avendano}}, \bibinfo{author}{\bibfnamefont{C.}~\bibnamefont{Bambi}}, \bibinfo{author}{\bibfnamefont{C.~A.~R.} \bibnamefont{Herdeiro}}, \bibnamefont{and} \bibinfo{author}{\bibfnamefont{E.}~\bibnamefont{Radu}}, \bibinfo{journal}{JCAP} \textbf{\bibinfo{volume}{07}}, \bibinfo{pages}{049} (\bibinfo{year}{2016}), \eprint{1606.04654}.

\bibitem[{\citenamefont{Cao et~al.}(2016)\citenamefont{Cao, Cardenas-Avendano, Zhou, Bambi, Herdeiro, and Radu}}]{Cao:2016zbh}
\bibinfo{author}{\bibfnamefont{Z.}~\bibnamefont{Cao}}, \bibinfo{author}{\bibfnamefont{A.}~\bibnamefont{Cardenas-Avendano}}, \bibinfo{author}{\bibfnamefont{M.}~\bibnamefont{Zhou}}, \bibinfo{author}{\bibfnamefont{C.}~\bibnamefont{Bambi}}, \bibinfo{author}{\bibfnamefont{C.~A.~R.} \bibnamefont{Herdeiro}}, \bibnamefont{and} \bibinfo{author}{\bibfnamefont{E.}~\bibnamefont{Radu}}, \bibinfo{journal}{JCAP} \textbf{\bibinfo{volume}{10}}, \bibinfo{pages}{003} (\bibinfo{year}{2016}), \eprint{1609.00901}.

\bibitem[{\citenamefont{Fabian et~al.}(2000)\citenamefont{Fabian, Iwasawa, Reynolds, and Young}}]{Fabian:2000nu}
\bibinfo{author}{\bibfnamefont{A.~C.} \bibnamefont{Fabian}}, \bibinfo{author}{\bibfnamefont{K.}~\bibnamefont{Iwasawa}}, \bibinfo{author}{\bibfnamefont{C.~S.} \bibnamefont{Reynolds}}, \bibnamefont{and} \bibinfo{author}{\bibfnamefont{A.~J.} \bibnamefont{Young}}, \bibinfo{journal}{Publ. Astron. Soc. Pac.} \textbf{\bibinfo{volume}{112}}, \bibinfo{pages}{1145} (\bibinfo{year}{2000}), \eprint{astro-ph/0004366}.

\bibitem[{\citenamefont{{Fabian} et~al.}(1989)\citenamefont{{Fabian}, {Rees}, {Stella}, and {White}}}]{1989MNRAS.238..729F}
\bibinfo{author}{\bibfnamefont{A.~C.} \bibnamefont{{Fabian}}}, \bibinfo{author}{\bibfnamefont{M.~J.} \bibnamefont{{Rees}}}, \bibinfo{author}{\bibfnamefont{L.}~\bibnamefont{{Stella}}}, \bibnamefont{and} \bibinfo{author}{\bibfnamefont{N.~E.} \bibnamefont{{White}}}, \bibinfo{journal}{Mon. Not. Roy. Astron. Soc.} \textbf{\bibinfo{volume}{238}}, \bibinfo{pages}{729} (\bibinfo{year}{1989}).

\bibitem[{\citenamefont{Tanaka et~al.}(1995)\citenamefont{Tanaka, Nandra, Fabian, Inoue, Otani, Dotani, Hayashida, Iwasawa, Kii, Kunieda et~al.}}]{Tanaka1995GravitationallyRE}
\bibinfo{author}{\bibfnamefont{Y.}~\bibnamefont{Tanaka}}, \bibinfo{author}{\bibfnamefont{K.}~\bibnamefont{Nandra}}, \bibinfo{author}{\bibfnamefont{A.~C.} \bibnamefont{Fabian}}, \bibinfo{author}{\bibfnamefont{H.}~\bibnamefont{Inoue}}, \bibinfo{author}{\bibfnamefont{C.}~\bibnamefont{Otani}}, \bibinfo{author}{\bibfnamefont{T.}~\bibnamefont{Dotani}}, \bibinfo{author}{\bibfnamefont{K.}~\bibnamefont{Hayashida}}, \bibinfo{author}{\bibfnamefont{K.}~\bibnamefont{Iwasawa}}, \bibinfo{author}{\bibfnamefont{T.}~\bibnamefont{Kii}}, \bibinfo{author}{\bibfnamefont{H.}~\bibnamefont{Kunieda}}, \bibnamefont{et~al.}, \bibinfo{journal}{Nature} \textbf{\bibinfo{volume}{375}}, \bibinfo{pages}{659} (\bibinfo{year}{1995}), \urlprefix\url{https://api.semanticscholar.org/CorpusID:4348405}.

\bibitem[{\citenamefont{Narayan}(2005)}]{Narayan:2005ie}
\bibinfo{author}{\bibfnamefont{R.}~\bibnamefont{Narayan}}, \bibinfo{journal}{New J. Phys.} \textbf{\bibinfo{volume}{7}}, \bibinfo{pages}{199} (\bibinfo{year}{2005}), \eprint{gr-qc/0506078}.

\bibitem[{\citenamefont{Dauser et~al.}(2013)\citenamefont{Dauser, Garcia, Wilms, Bock, Brenneman, Falanga, Fukumura, and Reynolds}}]{Dauser:2013xv}
\bibinfo{author}{\bibfnamefont{T.}~\bibnamefont{Dauser}}, \bibinfo{author}{\bibfnamefont{J.}~\bibnamefont{Garcia}}, \bibinfo{author}{\bibfnamefont{J.}~\bibnamefont{Wilms}}, \bibinfo{author}{\bibfnamefont{M.}~\bibnamefont{Bock}}, \bibinfo{author}{\bibfnamefont{L.~W.} \bibnamefont{Brenneman}}, \bibinfo{author}{\bibfnamefont{M.}~\bibnamefont{Falanga}}, \bibinfo{author}{\bibfnamefont{K.}~\bibnamefont{Fukumura}}, \bibnamefont{and} \bibinfo{author}{\bibfnamefont{C.~S.} \bibnamefont{Reynolds}}, \bibinfo{journal}{Mon. Not. Roy. Astron. Soc.} \textbf{\bibinfo{volume}{430}}, \bibinfo{pages}{1694} (\bibinfo{year}{2013}), \eprint{1301.4922}.

\bibitem[{\citenamefont{Chainakun et~al.}(2022)\citenamefont{Chainakun, Watcharangkool, and Young}}]{Chainakun:2022gcz}
\bibinfo{author}{\bibfnamefont{P.}~\bibnamefont{Chainakun}}, \bibinfo{author}{\bibfnamefont{A.}~\bibnamefont{Watcharangkool}}, \bibnamefont{and} \bibinfo{author}{\bibfnamefont{A.~J.} \bibnamefont{Young}}, \bibinfo{journal}{Mon. Not. Roy. Astron. Soc.} \textbf{\bibinfo{volume}{512}}, \bibinfo{pages}{728} (\bibinfo{year}{2022}), \eprint{2202.03657}.

\bibitem[{\citenamefont{Wilkins and Fabian}(2012)}]{Wilkins_2012}
\bibinfo{author}{\bibfnamefont{D.~R.} \bibnamefont{Wilkins}} \bibnamefont{and} \bibinfo{author}{\bibfnamefont{A.~C.} \bibnamefont{Fabian}}, \bibinfo{journal}{Monthly Notices of the Royal Astronomical Society} \textbf{\bibinfo{volume}{424}}, \bibinfo{pages}{1284–1296} (\bibinfo{year}{2012}), ISSN \bibinfo{issn}{0035-8711}, \urlprefix\url{http://dx.doi.org/10.1111/j.1365-2966.2012.21308.x}.

\bibitem[{\citenamefont{Baker and Young}(2022)}]{gradus}
\bibinfo{author}{\bibfnamefont{F.}~\bibnamefont{Baker}} \bibnamefont{and} \bibinfo{author}{\bibfnamefont{A.}~\bibnamefont{Young}}, \emph{\bibinfo{title}{Gradus.jl}} (\bibinfo{year}{2022}), \urlprefix\url{https://doi.org/10.5281/zenodo.6471796}.

\bibitem[{\citenamefont{{Cunningham}}(1975)}]{1975ApJ...202..788C}
\bibinfo{author}{\bibfnamefont{C.~T.} \bibnamefont{{Cunningham}}}, \bibinfo{journal}{\apj} \textbf{\bibinfo{volume}{202}}, \bibinfo{pages}{788} (\bibinfo{year}{1975}).

\bibitem[{\citenamefont{Bambi}(2012{\natexlab{b}})}]{Bambi:2012tg}
\bibinfo{author}{\bibfnamefont{C.}~\bibnamefont{Bambi}}, \bibinfo{journal}{Astrophys. J.} \textbf{\bibinfo{volume}{761}}, \bibinfo{pages}{174} (\bibinfo{year}{2012}{\natexlab{b}}), \eprint{1210.5679}.

\bibitem[{\citenamefont{{Laor}}(1991)}]{Laor1991}
\bibinfo{author}{\bibfnamefont{A.}~\bibnamefont{{Laor}}}, \bibinfo{journal}{\apj} \textbf{\bibinfo{volume}{376}}, \bibinfo{pages}{90} (\bibinfo{year}{1991}).

\bibitem[{\citenamefont{Pariev and Bromley}(1998)}]{Pariev:1998ei}
\bibinfo{author}{\bibfnamefont{V.~I.} \bibnamefont{Pariev}} \bibnamefont{and} \bibinfo{author}{\bibfnamefont{B.~C.} \bibnamefont{Bromley}}, \bibinfo{journal}{Astrophys. J.} \textbf{\bibinfo{volume}{508}}, \bibinfo{pages}{590} (\bibinfo{year}{1998}), \eprint{astro-ph/9806134}.

\bibitem[{\citenamefont{Chen and Pu}(2024)}]{Chen:2024hpw}
\bibinfo{author}{\bibfnamefont{C.-Y.} \bibnamefont{Chen}} \bibnamefont{and} \bibinfo{author}{\bibfnamefont{H.-Y.} \bibnamefont{Pu}}, \bibinfo{journal}{JCAP} \textbf{\bibinfo{volume}{09}}, \bibinfo{pages}{043} (\bibinfo{year}{2024}), \eprint{2404.07055}.

\bibitem[{\citenamefont{Bromley et~al.}(1997)\citenamefont{Bromley, Chen, and Miller}}]{Bromley:1996wb}
\bibinfo{author}{\bibfnamefont{B.~C.} \bibnamefont{Bromley}}, \bibinfo{author}{\bibfnamefont{K.}~\bibnamefont{Chen}}, \bibnamefont{and} \bibinfo{author}{\bibfnamefont{W.~A.} \bibnamefont{Miller}}, \bibinfo{journal}{Astrophys. J.} \textbf{\bibinfo{volume}{475}}, \bibinfo{pages}{57} (\bibinfo{year}{1997}), \eprint{astro-ph/9601106}.

\bibitem[{\citenamefont{Wu and Wang}(2007)}]{Wu:2007bq}
\bibinfo{author}{\bibfnamefont{S.-M.} \bibnamefont{Wu}} \bibnamefont{and} \bibinfo{author}{\bibfnamefont{T.-G.} \bibnamefont{Wang}}, \bibinfo{journal}{Mon. Not. Roy. Astron. Soc.} \textbf{\bibinfo{volume}{378}}, \bibinfo{pages}{841} (\bibinfo{year}{2007}), \eprint{0705.1796}.

\bibitem[{\citenamefont{Ori}(1991)}]{Ori:1991zz}
\bibinfo{author}{\bibfnamefont{A.}~\bibnamefont{Ori}}, \bibinfo{journal}{Phys. Rev. Lett.} \textbf{\bibinfo{volume}{67}}, \bibinfo{pages}{789} (\bibinfo{year}{1991}).

\bibitem[{\citenamefont{Poisson and Israel}(1989)}]{Poisson:1989zz}
\bibinfo{author}{\bibfnamefont{E.}~\bibnamefont{Poisson}} \bibnamefont{and} \bibinfo{author}{\bibfnamefont{W.}~\bibnamefont{Israel}}, \bibinfo{journal}{Phys. Rev. Lett.} \textbf{\bibinfo{volume}{63}}, \bibinfo{pages}{1663} (\bibinfo{year}{1989}).

\bibitem[{\citenamefont{De~Felice and Tsujikawa}(2025)}]{DeFelice:2024seu}
\bibinfo{author}{\bibfnamefont{A.}~\bibnamefont{De~Felice}} \bibnamefont{and} \bibinfo{author}{\bibfnamefont{S.}~\bibnamefont{Tsujikawa}}, \bibinfo{journal}{Phys. Rev. Lett.} \textbf{\bibinfo{volume}{134}}, \bibinfo{pages}{081401} (\bibinfo{year}{2025}), \eprint{2410.00314}.

\bibitem[{\citenamefont{Penrose}(1969)}]{Penrose:1969pc}
\bibinfo{author}{\bibfnamefont{R.}~\bibnamefont{Penrose}}, \bibinfo{journal}{Riv. Nuovo Cim.} \textbf{\bibinfo{volume}{1}}, \bibinfo{pages}{252} (\bibinfo{year}{1969}).

\bibitem[{\citenamefont{Nakao}(2024)}]{Nakao2024}
\bibinfo{author}{\bibfnamefont{K.-i.} \bibnamefont{Nakao}}, \emph{\bibinfo{title}{Revisit Cosmic Censorship and Black Holes}} (\bibinfo{publisher}{Springer Nature Singapore}, \bibinfo{address}{Singapore}, \bibinfo{year}{2024}), pp. \bibinfo{pages}{305--336}, ISBN \bibinfo{isbn}{978-981-97-1172-7}, \urlprefix\url{https://doi.org/10.1007/978-981-97-1172-7_11}.

\bibitem[{\citenamefont{Brenneman and Reynolds}(2006)}]{Brenneman:2006hw}
\bibinfo{author}{\bibfnamefont{L.~W.} \bibnamefont{Brenneman}} \bibnamefont{and} \bibinfo{author}{\bibfnamefont{C.~S.} \bibnamefont{Reynolds}}, \bibinfo{journal}{Astrophys. J.} \textbf{\bibinfo{volume}{652}}, \bibinfo{pages}{1028} (\bibinfo{year}{2006}), \eprint{astro-ph/0608502}.

\bibitem[{\citenamefont{Tiwari et~al.}(2018)\citenamefont{Tiwari, Fairhurst, and Hannam}}]{Tiwari:2018qch}
\bibinfo{author}{\bibfnamefont{V.}~\bibnamefont{Tiwari}}, \bibinfo{author}{\bibfnamefont{S.}~\bibnamefont{Fairhurst}}, \bibnamefont{and} \bibinfo{author}{\bibfnamefont{M.}~\bibnamefont{Hannam}}, \bibinfo{journal}{Astrophys. J.} \textbf{\bibinfo{volume}{868}}, \bibinfo{pages}{140} (\bibinfo{year}{2018}), \eprint{1809.01401}.

\bibitem[{\citenamefont{Li and Bambi}(2014)}]{Li:2013jra}
\bibinfo{author}{\bibfnamefont{Z.}~\bibnamefont{Li}} \bibnamefont{and} \bibinfo{author}{\bibfnamefont{C.}~\bibnamefont{Bambi}}, \bibinfo{journal}{JCAP} \textbf{\bibinfo{volume}{01}}, \bibinfo{pages}{041} (\bibinfo{year}{2014}), \eprint{1309.1606}.

\bibitem[{\citenamefont{Bambi}(2012{\natexlab{c}})}]{Bambi:2012ku}
\bibinfo{author}{\bibfnamefont{C.}~\bibnamefont{Bambi}}, \bibinfo{journal}{Phys. Rev. D} \textbf{\bibinfo{volume}{85}}, \bibinfo{pages}{043002} (\bibinfo{year}{2012}{\natexlab{c}}), \eprint{1201.1638}.

\bibitem[{\citenamefont{Bambi}(2012{\natexlab{d}})}]{Bambi:2012zg}
\bibinfo{author}{\bibfnamefont{C.}~\bibnamefont{Bambi}}, \bibinfo{journal}{Phys. Rev. D} \textbf{\bibinfo{volume}{86}}, \bibinfo{pages}{123013} (\bibinfo{year}{2012}{\natexlab{d}}), \eprint{1204.6395}.

\bibitem[{\citenamefont{Abdikamalov et~al.}(2019)\citenamefont{Abdikamalov, Abdujabbarov, Ayzenberg, Malafarina, Bambi, and Ahmedov}}]{Abdikamalov:2019ztb}
\bibinfo{author}{\bibfnamefont{A.~B.} \bibnamefont{Abdikamalov}}, \bibinfo{author}{\bibfnamefont{A.~A.} \bibnamefont{Abdujabbarov}}, \bibinfo{author}{\bibfnamefont{D.}~\bibnamefont{Ayzenberg}}, \bibinfo{author}{\bibfnamefont{D.}~\bibnamefont{Malafarina}}, \bibinfo{author}{\bibfnamefont{C.}~\bibnamefont{Bambi}}, \bibnamefont{and} \bibinfo{author}{\bibfnamefont{B.}~\bibnamefont{Ahmedov}}, \bibinfo{journal}{Phys. Rev. D} \textbf{\bibinfo{volume}{100}}, \bibinfo{pages}{024014} (\bibinfo{year}{2019}), \eprint{1904.06207}.

\end{thebibliography}

\end{document}